# Towards a Democratic University:
# A call for Reflexive Evaluation and a Participative Culture

**Julia Heuritsch**
Research Group "Reflexive Metrics", Institut für Sozialwissenschaften, Humboldt Universität zu Berlin, Universitätsstraße 3b, 10117 Berlin, Germany; julia.heuritsch@hu-berlin.de & julia.heuritsch@gmail.com

## Abstract

The extensive focus on performance indicators in research evaluation has been facing critique in science studies. Stemming from a neoliberalist paradigm, metrics allegedly objectify and create certainty about researchers' performance. This has created a publish-or-perish culture where deviant behaviour, such as research misconduct, may have become the rule of the game. Not only does this culture foster a decrease of scientists' well-being, but also a decrease in research quality. In recent years, calls for a culture change have accumulated, from discussing detrimental cultural aspects under #IchbinHannah to studies that demonstrate the connection between research culture and research integrity. This study is a qualitative analysis of how astronomers reimagine their research culture. This includes alternative output formats, alternative evaluation criteria and how they aspire to do research differently. In summary, we find that the time is ripe for a transformation in research culture towards a more participative and diverse work environment. This may include the use of an open knowledge management infrastructure, where all sorts of research output of various stages can be stored and shared. Moreover, through a more reflexive evaluation, which is continuously adapted to the needs of the researchers, scientific quality may be encouraged, rather than producing more and more publications. This study sets the basis for future action research with the aim of transforming academic cultures towards more participative ones, where scientists can let their creative minds thrive and collaborate beyond disciplinary boundaries.

1. ## Introduction

Ideally, science is performed by associations of people, having an "authoritarian-free discourse" (Eitel & Mlynek, 2014: 33; cf. Habermas,1968). This ideal roots back to the medieval university, which "has had a dominant position in modernized versions until about 1980" (Bleiklie, 2018: 2). On the one hand, autonomy has always been important to be creative and perform good research, on the other hand people, who work towards the same goal, need some kind of leadership and organisational structures (Eitel & Mlynek, 2014). This tension between autonomy and leadership is amplified by the fact that in recent decades, the institution of a university has become ever more market-like under the neoliberalist paradigm (Lorenz, 2012). Researchers need to be accountable towards taxpayers, politicians and other stakeholders. In order to receive funding and proceed in their careers, they face fierce competition, wherein quantitative indicators measure their performance (e.g. Anderson et al., 2007; Moosa, 2018). That is why the focus of research evaluation has been on accountability of the individual.

Evaluating performance based on solely a few quantitative indicators, serving as a proxy for the quality of work that researchers produced, has faced critique in recent years (for an overview see Fochler & De Rijcke, 2017). In socio-technical systems, performance measures do not only describe, but also prescribe behaviour (Desrosières, 1998). Instead of working towards a goal out of autonomous motivation (identification with the action), people find themselves acting out of controlled motivation (the need to hit the target; Heuritsch, 2021c)[1]. Acting out of controlled motivation to hit a target tempts people to adopt innovative gaming strategies to hit the aspired target and therefore bereaves the measure that is targeted of its meaning (Goodhart's law[2]). Some studies report that gaming, including research misconduct, already constitute the norm in science (e.g. De Vries et al., 2006; John et al. 2012). Hence, the "bad apple narrative" has been rendered as too simplistic to explain behaviour which threatens research integrity (Haven & Woudenberg, 2021). In other words, while individual dispositions may increase the likelihood of research misbehaviour, the structural conditions and culture of an organisation, such as the scientific system, are much more predictive in shaping researcher's behaviour on a large scale (e.g. Crain et al., 2013; Martinson et al., 2013 & 2016; Wells et al., 2014). For example, if members of an organisation feel treated fairly in terms of research allocation (distributive justice) and how processes are handled (procedural justice), they are more likely to respond with favourable behaviour (Martinson et al., 2006). With the increasing dependence on "external resources, whether from government or private industry" (ibid.: 3), and undergoing neoliberalist-inspired reforms to stay competitive for these resources, the university has changed from a "republic of scholars" to a "corporate enterprise" (Bleiklie, 2018: 2). Since the resulting research culture seems to foster the normalisation of scientific misconduct and a decrease in research integrity, it is time to scrutinise this culture, rather than the individual.

Action research is a social science methodology which is an interactive inquiry, aimed at understanding organisational culture and -processes and transforming them at the same time. In contrast to conventional social science methods, which aim at studying social settings without interfering in order to generate "objective" knowledge (Fricke, 2014: 222), action research is inherently participative and normative. It is participative in the sense that employees of the organisation under scrutiny are involved in many steps of the research, which may range from selecting the methods to analysis and learning from the findings. Action research inspires a collective (self-) reflection on otherwise tacit and unquestioned organisational norms, values, processes and structures. It is normative in the sense that action research is based on democratic values[3]. The aim is the development of democratic structures and fostering democratic behaviour, by facilitating participative dialogues (ibid.: 215). Action research is a reflexive endeavour in that – through reflection together with those being studied – it shapes the practices and perspectives of everybody involved. That is how through action research organisations may transform towards more democratic institutions.

Box 1 – Deliberative & participative democracy

Most democratic nations employ a representative democracy, where representatives are elected and entrusted to carry out the business of governance. These representatives form the government. Dahler-Larsen (2019) refers to this type of democracy as the *government model* of democratic systems.

As opposed to representative democracy, in a deliberative & participative democracy, all members of the democratic institution may directly participate in decision making processes through participative procedures, such as dialogue (e.g. Pateman, 1970). Dahler-Larsen (2019) refers to this type of democracy as the *governance model* of democratic systems.
Deliberation and pluralism are key democratic principles, and even more important in a participative democracy (e.g. Habermas, 1968; Dahl, 2006; Sunstein, 2018). The legitimate diversity of opinions and interests, resulting in debates and controversy, is democracy's biggest challenge at the same time (e.g. Foroutan, 2019). Especially in a world, which becomes increasingly **v**olatile, **u**ncertain, **c**omplex and **a**mbiguous (VUCA[4]), a democratically competent agent is required to have *ambiguity tolerance* (Habermas, 1968: 128). This refers to the acceptance that in a (socially) complex world there may not just be an either/or, but a plurality of seemingly contradicting perspectives.

---

[1] In short, autonomous motivation comes from identifying with the value of the activity, while controlled motivation is instrumental to acquire or avoid external rewards or punishments (Gagné et al., 2010).
[2] "When a measure becomes a target, it ceases to be a good measure." (https://en.wikipedia.org/wiki/Goodhart%27s_law)
[3] For our understanding of "democracy" see *Box 1*.

As compared to the government model of democratic systems, involving hard laws, the modes of regulation in the governance model are soft laws. These involve guidelines, recommendations, agreements, indicators, standards, and benchmarks, none of which are binding (Dahler-Larsen, 2019: 191). Governance networks may therefore be more dynamic in navigating in complex environments, while depending on the active consent of their participants and their ongoing deliberation (ibid.: 191).

The *evaluative inquiry* (EI[5]) may be understood as an application of action research (e.g. Coghlan & Brydon-Miller, 2014). Researchers from the Center for Science and Technology studies (CWTS) have adapted the EI to the context of the evaluation culture in academia and describes the methodology in a series of blog articles[6]. While "mainstream evaluation metrics tend to understand academic value as performance", the EI also investigates "other valuable elements of academic value trajectories"[6]. It does so by shifting the focus from "talking about" to "thinking with"[7] scientists and stakeholders and thereby employing a combination quantitative and qualitative methods, depending on what fits the specific evaluation purpose best. The EI is not only a *participatory* approach to evaluation, but also a *reflexive* one. It is so in two ways: First, those conducting an EI reflect together with those involved in a respective evaluation on the context they are producing knowledge and what they value about their work (output). The EI "supports academic units in crystalizing goals, missions, visions and strategy, taking stock of the diversity of output, making the multitude of stakeholder relations visible, and listing staff opinions about the academic organization."[6]. Hence, finding a common purpose[8] and narrative on how the organisation performs research and organises itself, is of crucial value of this step. Second, the EI promotes a continuously ongoing reflection on the reflexiveness of evaluation procedures (i.e. their feedback effects). The evaluative inquiry therefore "values both, the processes and the outcomes of evaluation" (Coghlan & Brydon-Miller, 2014: 1) and delivers valuable learning opportunities for everybody involved.

---

[4] https://www.vuca-world.org/
[5] On a side note: In participation research, the abbreviation EI actually denotes "employee involvement", which is a convenient coincidence.
[6] Series of blog articles written by members of the Centre for Science and Technology Studies (CWTS) on the Evaluative Inquiry:
https://www.leidenmadtrics.nl/articles/evaluative-inquiry-i-academic-value-is-more-than-performance
https://www.leidenmadtrics.nl/articles/evaluative-inquiry-ii-evaluating-research-in-context
https://www.leidenmadtrics.nl/articles/evaluative-inquiry-iii-mixing-methods-for-evaluating-research
https://www.leidenmadtrics.nl/articles/evaluative-inquiry-iv-accountability-and-learning
https://blogs.lse.ac.uk/impactofsocialsciences/2018/11/29/the-evaluative-inquiry-a-new-approach-to-research-evaluation/
[7] Presentation by Maximilian Fochler at STS Conference Graz (AUT) 2019 on 6th May 2019
[8] Purpose is a concept, which is often associated with a transformation of organisational culture towards a more participative one. A systematic account of its history and rich meaning is beyond the scope of this research. The way we use the term purpose, it refers to the meaningful goal(s) an organisation aspires to achieve. If this organisational purpose is aligned with the employees' individual goals, employees may work with more engagement and responsibility. This is because their autonomous motivation may be fostered by doing something that is meaningful and valuable to them (e.g. Schnell, 2018).

> "Evaluative inquiry is a way of fostering individual and team learning within an organization about issues that are critical to its purpose and what it values. It involves an intentional process of framing important questions, systematically gathering information relevant to the questions, and using the information to draw credible conclusions that can shape practice." (Parsons, 2009)

The study of feedback effects of the use of metrics in performance measurement on the research behaviour, -culture, knowledge production processes and research quality is called *reflexive metrics*. Reflexive metrics is a branch of science studies and comprises of the sociology of (e)valuation and the sociology of quantification. There have been a series of studies under the umbrella of reflexive metrics which may set the basis for an EI and a transformation of research culture in astrophysics. Heuritsch (2019) investigated what astronomers themselves value as good quality research practices and output. The author found three quality criteria: (1) good research needs to push knowledge forward, which includes studying a diversity of topics and making incremental contributions; (2) The research needs to be based on clear, verifiable and sound methodology that is (3) reported in an understandable and transparent way. This includes the sharing of data and reduction code.

In a follow-up study, Heuritsch (2021a) studied the "organisational hinterland" (Dahler-Larsen, 2019) of astronomy to understand what performance indicators measure, how they diverge from the astronomers' definition of quality and how this discrepancy affects research behaviour. In a nutshell, Heuritsch (2021a) finds evidence for the over-emphasis on performance measured by publication rate, reception of external grants and telescope time, leading to gaming strategies to score well on those indicators. These are found to be a response to the dissonance between cultural values (producing qualitative research that genuinely pushes knowledge forward) and the institutional objectives imposed to have career in academia (scoring well on indicators). In other words, there is a discrepancy between what indicators measure and the astronomers' definition of scientific quality – the so-called evaluation gap. Gaming strategies then give the appearance of compliance with cultural values, while using institutionalised means to achieve a good bibliometric record in innovative ways, such as salami slicing, cutting corners or going for easy publications. The author finds evidence for astronomers prioritizing quantity over quality of publications.

Both studies (Heuritsch 2019 & 2021a) were conducted by means of qualitative interviews. Following up on Heuritsch (2021a), Heuritsch (2021b) performed a quantitative study, surveying international astronomers worldwide. The author found that publication pressure explains 10% of the variance in occurrence of misconduct and between 7 and 13% of the variance of the perception of distributive & organisational justice as well as overcommitment to work. Further, the survey showed that the epistemic harm of questionable research practices on research quality should not be underestimated.

The finding that publication pressure decreases the perception of a fair organizational culture and increases the likelihood of scientific misconduct stems from the fact, that "printed publications have been considered the sole publishing objective of research for centuries" (Breuer & Trilcke, 2021: 5). We therefore consider a meta-reflection on publication formats as a vital aspect for an evaluative inquiry. A shift of a focus away from the traditional format of publication is already under way due to digital technologies (ibid.). It seems vital to include researchers themselves to investigate how "digital technologies can support and change the growth of scientific knowledge, its reproducibility and the accessibility of said knowledge" (ibid.: 4). Such an endeavour aims at the development of context-/ discipline specific formulation of "clear criteria for the *recognition of different digital publication formats as* *attributable and remunerable scientific practice of the type 'publishing'"* (ibid.: 4; italics in the original).

The aim of this study is to reimagine cultural aspects of research in astronomy, such as publication formats and ways of being assessed. While previous literature has pointed to structural problems in academia and how they affect knowledge production processes and research integrity, the search for contextual solutions is a novel approach. Building up on these studies, especially the ones conducted about the field of astronomy, this study is based on qualitative interviews and open survey questions. The goal is to understand astronomers' perspectives on cultural changes that are desired and/or already under way and to work out specific recommendations about how to support these. These may provide a basis for any future action researchers, working directly together with astronomers to transform research culture into a democratic and flourishing one.

This paper is structured as follows: First, in *section 2* we introduce studies regarding the status quo of academic culture, which operates in a neoliberalist paradigm entailing metrics as performance measures. This section also includes an overview of recent calls for a culture change and what that could entail. The method section refers to the sample selection & procedure and describes the (interview) questions this study is based on. The result section is structured according to the three "miracle questions" asked to our study participants: reimagining alternative output formats, evaluation criteria and research. The discussion section introduces recommendations, whose implementation may lead to a more participative research culture, based on this study's results, literature and existing initiatives. Finally, we draw our conclusions, implications thereof and give an outlook for future research in *section 6*.

## 2. The research culture: It's status quo and where to go

### 2.1. Indicators as Quality inscriptions

New Public Management (NPM) is a "reform model arguing that the quality and efficiency of the civil service should be improved by introducing management techniques and practices drawn mainly from the private sector" (Bleiklie, 2018: 1). NPM works under the neoliberalist paradigm, which is an "ideology and policy model that emphasizes the value of free market competition" (ibid.: 1). With increased dependence on external funding, whether from government or from the private sector, academic institutions have undergone neoliberalist reforms since the 1980s (Lorenz, 2012). In order to serve the interests of stakeholders, such as funders and tax payers, hierarchical structures have been put into place "to provide leaders with authority and managerial resources to make and enforce strategic decisions within the organization" (Bleiklie, 2018: 2). Today, most of our organisations – whether private or governmental – are exposed to the "tyranny of the market" due to neoliberalism being the prevailing ideology in policy making (Moosa, 2018: 15). Academic freedom has been increasingly circumscribed by the organisational structures, resulting from this operating system[9].

The neoliberalist understanding of efficiency calls for measuring individual's performance by quantitative indicators. While intended as proxies for qualitative work output, performance indicators resemble *quality inscriptions* (Dahler-Larsen, 2019). A quality inscription is a "documentation of quality, usually in the form of quantification" (ibid.: 19). Quality inscriptions enable allegedly objective comparisons, ratings and rankings. That way conceptualising quality has an important social function: it helps structuring reality. However, because quality itself is a social construct, which depends on time and context, quality does not merely "'indicate' and underlying objective reality", but "defines reality in a particular way" (ibid.: 10). That way quality inscriptions have constitutive effects on social reality and actors' behaviour: "Once the quality of a phenomenon is expressed through an indicator, practices turn toward what is seen as important in the light of the indicator" (ibid.: 143). This is the basis for Goodhart's law and the reason for a shift from autonomous to controlled motivation. By fixing quality by a few numerical indicators, professional values, especially more tacit and context-dependent ones, may be undermined (ibid.).

---

[9] Term neoliberalism as "operating system" is taken from: https://www.sueddeutsche.de/kultur/peter-andre-alt-sachbuch-1.5444881

Because the notion of quality is inherently positive and quantitative quality inscriptions in form metrics give the impression of objectivity, quality indicators are desired neoliberal instruments (Dahler-Larsen, 2019). Quality inscriptions therefore are powerful rhetorical tools, often imposed by decision makers (such as stakeholders on top of the hierarchy), rather than those affected by the decisions and, given the positive notion of quality, difficult to question by the ones affected: "the measurement arrives as a *fait accompli* rather than an object for discussion" (ibid.: 202; italics in the original).

> "In order to secure commensuration and transferability, qualitization cuts public matters into small pieces that are compared and assessed, traded and exchanged. Quality not only promotes a neoliberal agenda; it helps extend it into domains that monetarization has yet to penetrate". (Dahler-Larsen, 2019: 209)

*2.2. Publish-or-Perish – a part of a neoliberalist academic culture*

Arguably, the most prominent quality inscription in academia is a scientists' publication rate. This is because "the 'paper' is, for historical reasons, a holy cow" in terms of what counts as an output (Albrecht, 2007: 144). The number of first author publications is crucially important for a researcher's career. This gives rise for the prevailing "publish or perish" (POP; Moosa, 2018) culture in academia, which is deeply intertwined with the neoliberalist reforms of the academic institution. The perceived benefits of the "publish or perish" imperative, from an NPM perspective, is that it generates a pressure to publish which supposedly separates the wheat from the chaff, letting only the most productive scientists – those who return the biggest value to their stakeholders as measured by the quality indicator – survive on the career ladder. This causes a rat race to produce ever more research at the expense of qualitative aspects that are not being measured; such as replicability, readability, comprehensiveness and value for the research community. Moreover, there is no merit in non-article publications, unpublishable negative results and non-research activities, such as teaching, community service and acting on one's research findings (ibid.).

> "Reporting of the results of scientific research must be governed by the principle of telling the truth, the whole truth and nothing but the truth." (Moosa, 2018: 66)

Under POP, however, the researcher's motivation shifts from an autonomous one of discovering truths to a controlled one of generating publications. Publishing has become an end in itself, rather than the means to an end (ibid.). Under the pressure to hit targets not only researcher's physical and mental well-being are decreasing, but also a tendency to indulge in research misconduct in all shapes and forms has arisen (Tijdink, 2014; Moosa, 2018; Wellcome, 2020).

It may not be an exaggeration to worry that gaming (ranging from questionable research practices to outright fraud) may has become as necessary for a scientist to survive in academia as doping is for a cyclist to win the Tour de France. In a culture where ever more is what is aimed for, even the most honest among us will eventually cut some corners.

> "If university funding is determined by the rules of POP, then university management has no alternative but to oblige." (Moosa, 2018: 22)

> "This is chokepoint capitalism at its finest: publishers' primary 'asset' is a legally defensible barrier between academics and their career prospects, so it can coerce them into accepting all kinds of abusive conduct."[10]

### 2.3. Calls for a transformational culture change

Not only the POP culture arguably impedes the discovery process, but ever more voices criticizing the prevailing organisational culture in academia come to light. A survey titled "What Researchers Think About the Culture They Work In" (Wellcome, 2020) found that 78% of the researchers think that high levels of competition lead to an unkind working culture. While respondents accept competition as part of their job, they perceive the obsession with quantity of outputs, and narrow concepts of what counts as such, as a source of high pressure to meet metrics. 43% of the respondents perceive that their workplace puts more value on metrics than on research quality and 58% disagree with the statement that "current metrics have had a positive impact on research culture". Next to the perceived decrease in research quality (such as reproducibility and cherry-picking of results), long working hours, widespread concerns about job security and a decrease in physical and mental health are the consequences of this culture.

> "These results paint a shocking portrait of the research environment – and one we must all help change. A poor research culture ultimately leads to poor research. The pressures of working in research must be recognised and acted upon by all, from funders to leaders of research and to heads of universities and institutions."
> – Jeremy Farrar, Director of Wellcome (Wellcome, 2020: 3)

The importance of all involved parties, such as policy makers, university leadership, professors and supervisors, taking responsibility with respect to the prevailing research culture is also recognised by the discussions held under the hashtag #IchbinHanna[11] ("I am Hanna"). This hashtag first appeared in social media in Germany in 2021 to speak up about the precarious working conditions in academia, which are identified as the symptoms of an underlying unhealthy system. A system which not only involves the systematic exploitation of young researchers, but also fosters individualism – every researcher represents their own "academic me-inc."[12]. #IchbinHanna inspires the call for a more agile[13], resilient, transparent and

---

[10] https://pluralistic.net/2021/10/28/clintons-ghost/#cornucopia-concordance
[11] A few examples:
https://jungle.world/artikel/2021/27/alles-hochschule
https://www.jmwiarda.de/2021/09/01/hanna-helfen-geht-nur-gemeinsam/
https://www.faz.net/aktuell/feuilleton/debatten/hashtag-ichbinhanna-junge-akademiker-protestieren-17390890.html
https://www.sueddeutsche.de/kultur/universitaet-mittelbau-zeitvertrag-prekariat-promotion-habilitation-arbeitslosigkeit-1.5323946?reduced=true
[12] https://www.heise.de/tp/features/Die-Kritik-am-Wissenschaftszeitvertragsgesetz-ist-ein-Anfang-6120718.html
[13] Agile management methods originally stem from software development in the 1990s. The digital transformation brought a departure from linear process/ production models to more flexible and user-oriented ones (Boes et al., 2018). Despite the

participative culture instead[14]. The science journalist Jan-Martin Wiarda poses that "modern leadership in science means a professional leadership of science through itself" (ibid.). Many other discussions on (social) media[15] focus on the need of a transformation[16] of research culture, system change and reinvention of leadership in science. This development fits in with a general uneasiness regarding the adequacy of our current organisational models and management methods, slowly arising in the last decades.

> Box 2 – Paradigm changes in organisational models
>
> Laloux's "Reinvening Organisations"[17] is perhaps one of the most influential and forward-looking management books of the last decade. He colour-codes the different ways people have been organising themselves from the smallest tribes around 100.000 BC up until today. He describes how shifts in human consciousness change the way we structure and manage our organisations, because they are "simply the expression of our current world-view" (ibid.: 15).
>
> Since the age of enlightenment and industrial revolution, achievement-orange has been the prevailing paradigm. The breakthrough aspects, compared to the earlier paradigm, are innovation, accountability and meritocracy. In this paradigm, we have made big leaps towards (however, certainly not fully accomplished) social fairness and accumulated (material) wealth. As its name tells, what counts in this paradigm, is achieving profit and growth. Success is mostly measured in terms of money and recognition, or other quantitative quality inscriptions. By placing a carrot, shareholders motivate workers to work ever faster and more efficiently. This paradigm thinks of organisations as machines, where predict & control resemble the order of business.
>
> However, our life style has accelerated under the achievement-orange paradigm, which comes with an ever lager need to accept that we live in a VUCA world; The machine metaphor and linear thinking (such as the bad apple narrative) are too simplistic in a world that is inherently **v**olatile, **u**ncertain, **c**omplex and **a**mbiguous.

---

potential of agile methods to lead to a more participative and inclusive culture, these authors show that this relationship is ambiguous and not yet studied well.

[14] https://www.jmwiarda.de/2021/07/21/das-einzige-licht-am-ende-des-tunnels/
[15] See examples:
https://www.forschung-und-lehre.de/politik/fiktiver-name-in-zahlreiche-autorenlisten-geschmuggelt-3588/
https://media.nature.com/original/magazine-assets/d41586-020-01144-8/d41586-020-01144-8.pdf
https://www.nature.com/articles/d41586-021-01579-7
https://www.jmwiarda.de/2021/08/17/zur%C3%BCck-in-die-gesellschaft/
https://www.nature.com/articles/s41550-019-0961-2
https://www.nwo.nl/en/position-paper-room-everyones-talent
https://scholarlykitchen.sspnet.org/2020/02/18/reforming-research-assessment-a-tough-nut-to-crack/
https://www.tagesspiegel.de/wissen/riskante-forschung-mutige-ideen-scheitern-erlaubt/25214880.html
https://www.zeit.de/2021/09/streitkultur-universitaeten-wissensaustausch-no-platforming-wissenschaft
https://www.theguardian.com/education/2020/jan/15/universities-must-overhaul-the-toxic-working-culture-for-academic-researchers
https://www.duz.de/beitrag/!/id/1110/schon-immer-politisch
https://www.theguardian.com/society/2020/jan/15/researchers-facing-shocking-levels-of-stress-survey-reveals
https://www.youtube.com/watch?v=RLLoxxtBu50
https://www.digigebf21.de/frontend/index.php?folder_id=3857&page_id
https://www.zeit.de/campus/2021-10/ichbinhanna-hochschule-sabine-kunst-birgitt-riegraf-paderborn-befristete-stellen-mittelbau
[16] For a non-scientific, but, in our opinion, well-written overview of different paradigms of organisational models, see *Box 2*.
[17] Laloux, F. (2014), "Reinventing Organizations: A Guide to Creating Organizations Inspired by the Next Stage of Human Consciousness", Brussels: Nelson Parker.
For a short, but comprehensive summary see: https://pdf4pro.com/view/frederic-laloux-reinventing-organizations-5aab3c.html

In contrast, "evolutionary-teal" organisations accept VUCA and view organisations as living organisms. Concepts, such as hierarchy or power, are not abandoned but transcended, which means that they may be evoked depending on the context. Self-management is the order of business, which enables more flexible responses to change, rendering teal organisations more resilient. Moreover, they value working together towards a purpose and the health of the members of the living organism above all else, accepting non-quantitative and inter-subjective indicators as evidence.

While "Reinventing organisations" is evidence-based, it is not strict scientific research. There are certainly many parallels between what we refer to as "participative" in this paper and what Laloux calls "teal" leadership/ culture/ organisations. In what ways these concepts differ and overlap could be subject to further research.

Discussions among scientists from various disciplines during the symposium "Führen(d) in der Wissenschaft" ("Leading in science"; Eitel & Mlynek, 2014) held in Berlin in May 2014, underline the call for the current science system to be transformed into a more participative organisation. The global and complex organisation we call academia may benefit from leadership which supports self-organisation, (self-) reflection and motivation through purpose and praise. "Leadership is about visions and people"; It may draw on management's "facts and figures" (ibid.: 17), however does not perceive see numbers as the only way to describe reality. Participative leadership[18] in science has the potential to foster and support researchers in various ways needed and to facilitate enough room for creativity and academic freedom (ibid.: 57-58). The autonomous motivation of scientists to do research out of curiosity, which is arguably one of the biggest purposes of basic science, could be fostered by allowing for self-management and praise (ibid.: 107). A "healthy" research culture is one that allows for making mistakes, fostering strengths and guidance in weaknesses (ibid.: 44). This could foster innovation, the capacity to work in a team and self-organised informal or semiformal (interdisciplinary) collaborations (ibid.: 78). In such a culture flat hierarchies and a discourse on eye level could be the order of business. In the discussants' vision, leadership in science could sustain the ambivalence between facilitating structures and granting freedom. For example, the momentum of the self-organised system could be appreciated, while at the same time a culture of research quality is established (ibid.: 30). This may involve the development of evaluation criteria, that are, on the one hand, scientifically acceptable (evidence-based), and, on the other hand, not too rigid too allow for context and innovation. Such an academic organisational model of the future may already be "integrated in the DNA of a scientist" (Eitel & Mlynek, 2014: 120).

In a participative organisation, such evaluation criteria would neither be developed, nor implemented top-down, but in cooperation with the people involved – the researchers on "the shop floor". We believe that the *evaluative inquiry* could serve as a starting point to work out the narrative of the academic institution, with a focus on its purpose, together with the members of the institution. On that basis, also participatively, evaluation criteria may be developed, that would have the potential[19] to actually foster researchers' autonomous motivation to perform quality research.

---

[18] For more information on "participative leadership" refer to for example Rybnikova, I. & Lang, R. (2020), "Partizipative Führung: Auf den Spuren eines Konzeptes", *Gruppe. Interaktion. Organisation. Zeitschrift Für Angewandte Organisationspsychologie*, *51*(2), p.141–154. https://doi.org/https://doi.org/10.1007/s11612-020-00512-2

[19] The attentive reader with ambiguity tolerance will keep in mind that there is no guarantee for a participative culture to be more beneficial for its employees and with respect to achieving its goals than a strongly hierarchical one. Despite the literature introduced in this paper and the results of our own study, which highlight the need for more agile & participative ways of working together, we remind the reader that a transformation in organisational culture is to a large extent uncharted territory – both, in a scientific and in an explorative sense. Besides, usually advantages do not come without disadvantages.

*2.4. Democracy calls for Reflexive Evaluation: Evaluative Inquiry*

The key to a more democratic research culture, where both, researchers themselves and the quality of the work they produce can flourish, is to recognise the inherent self-organising dynamic of human organisations. An organisation is an emergent phenomenon, which means that it is more than the sum of its parts; not only the parts itself make up the system, but also the relationships among them (Parsons, 2009, Haken & Schiepek, 2006). What Parsons[20] (2009) refers to as *complex adaptive system theory* (CAS theory), is what Haken & Schiepek (2006) refer to as *synergetics*: an interdisciplinary system theory explaining the formation of patterns and structures in open systems, such as social ones.

> "In complex adaptive systems, many semi-independent and diverse agents, who are free to act in unpredictable ways, continually interact with each other. They are adapting to each other and the environment as a whole. They can create influential system-wide patterns." (Parsons, 2009: 48)

CAS theory and synergetics challenge how social systems are managed. Traditionally, organisational or institutional management has focussed on "maintaining stability" (Parsons, 2009: 48) by exercising top-down control. The machine metaphor prevailing in current organisations comes from our linear thinking (Morgan, 2018). We assume "that if the factors that influence a system are known, these factors can be controlled and outcomes predicted" (Parsons, 2009: 48). By contrast, synergetics demonstrates that complex systems are governed by chaos theory. In short, slightly different input conditions may create large differences in outcome, due to the complex interactions of the system parts (i.e. actors). Participative leadership, which accepts the self-organising character of social phenomena, such as organisations, is capable of understanding, encouraging and embracing their complexity. This type of leadership works with instead of against complexity and therefore is prone to lead more resilient organisations in a VUCA world. As such, participative leadership actually "*capitalises* on the self-organising dynamic" of organisations (ibid.: 49).

While participative leadership does not attempt taking control over the organisation, leveraging self-organisation does not mean to leave everything to unreflected unfolding. In fact, to support a self-organising dynamic, it is important to find guiding principles which are congruent with the purpose of the organisation and the basic values of those involved (Parsons, 2009). These guiding principles need to be continuously "renegotiated and calibrated through dialogue" (Dahler-Larsen, 2019: 191) given the dynamic characteristic of complex systems.

Evaluation as a practice is, in fact, rooted in a democratic policy as well (Dahler-Larsen, 2019: 189). The determination of official political goals "constitute the basis for evaluative criteria" (ibid.: 190). Since neoliberalism is ruling the political agenda of today, evaluation procedures are mainly based on quality inscriptions, which do not recognise the situation-bound and context-specific character of quality. These allegedly objective evaluation criteria travel rigidly through time and space, without being included in a continues deliberation process. This is already concerning from a democratic perspective on its own merit, but additionally evaluation criteria produce constitutive effects, which are usually not recognised by neither the evaluators, nor the evaluated. The term "constitute effects" was coined by Dahler-Larsen (2014) to describe "how a quality inscription influences reality while it claims to describe it" (Dahler-Larsen,

---

[20] Long before 'agile' ways of working have penetrated the management realm, Parsons actually developed the AGIL paradigm as an acronym reflecting on the four functional prerequisites an action system (incl. social systems) needs to entail to maintain self-preservation. The four functions are: Adaption – Goal Attainment – Integration – Latency.
(Parsons, T., 1951, "The Social System", Routledge, London, ISBN 0-415-06055-9)

2019: 117). Constitutive effects are naturally happening in complex, non-linear systems, and cannot be 'managed away', but need to be recognised as part of the culture and included in the deliberation process. Instead, in most of today's institutions and organisations constitutive effects escape deliberation, by moving unnoticed and "swiftly through time and space" (ibid.: 199). This undermines democratic legitimacy and "quality without democracy is concerning" (ibid.: 190) as it escapes the consent of people involved.

It may therefore be time for the academic institution, as a progressive organisation, to perform a more *reflexive evaluation*. A reflexive evaluation can be understood as the application of action research to the evaluation culture in academia. By developing evaluation criteria in a participatory and reflexive process, not only democratic values are pursued, but also constitutive effects of evaluation procedures are recognised. Reflexive evaluation would be a fluid and continuous negotiation. Constitutive effects could be leveraged for academia to grow a more participative research culture. This in turn would support reflexive evaluation and democratic participation. The capitalisation on self-organisation and its symptoms like constitutive effects, could therefore lead to a positive feedback-loop between research culture, evaluation and research quality.

The evaluative inquiry (EI) may be a good start, approaching a reflexive evaluation. The EI is a novel, exploratory and less standardised approach, designed to attend the "growing need in research organizations for interactive, formative and tailor-made evaluation services"[6]. The EI can be understood as a situated evaluation, which includes the perspectives of the researchers and "reveals epistemic commitments and community values"[21]. With its intention to revise the linear notion of impact[6] it is designed as a tool to handle the complexity that a VUCA world presents us with (Parsons, 2009).

CWTS researchers[6] have distilled four principles underpinning the evaluative inquiry's method: 1) the open-ended and relative concept of research value, 2) its contextualisation, 3) a mixed-method approach, 4) a focus on accountability *and* learning. The following paragraphs are dedicated to these principles.

First, the EI acknowledges that value, such as research quality cannot be fixed in form of qualitization without continues reflection about it. The EI proposes that "*knowledge diplomacy*"[6] is a crucial skill to negotiate what is of value. Instead of remaining in stagnation with respect to a mundane and bureaucratic academic evaluation, "the diplomat negotiates ways forward together, despite an apparent incongruence of worldviews or ambitions"[6]. This requires ambiguity tolerance[22]. Similarly, the EI does not argue against the use of performance metrics and accountability, but advocates for making other elementary aspects of academic value trajectories visible. As such, it *transcends* the tension arising from performance metrics not fully capturing concepts of value. The EI further transcends the need to distinguish between scientific and societal impact of research, since it is "interested in value as the connection between the institution's mission or research themes and their reception and use by others in the world around the institution"[6]. Hence, the EI's purpose is to support a research organisation to find its *purpose* and effective ways to work towards it.

Second, the EI acknowledges that research value is always situated in at least two contexts; the research organisation and the stakeholder – including the user (e.g. society) – context. Many factors make up the organisational context, such as "organizational histories, publication cultures, teaching versus research obligations, funding sources and epistemic cultures"[6]. All these difficult-to-quantify contextual factors influence knowledge production processes, what is relevant and what counts as scientific output. By taking these complex interactions into account, the EI moves away from putting the focus on individual excellence[6].

---

[21] Presentation by Tjitske Holtrop at the STI Conference 2018 in Leiden (NL) on 12th September 2018
[22] For a definition see *Box 1*.

Third, mixing quantitative and qualitative methods in complementary ways "provides different pieces to this complex puzzle, allowing for a less dichotomized and more contextualized approaches"[6]. An important part of the EI's methodological strategy is to work together with members of the academic institute, which includes participation in the analysis. Participation not only encourages involved members and parties to "rethink established ways of evaluating academic quality"[6], but also creates ownership of results. This may enhance the feeling of autonomy of those involved which is shown to be a predictor of work satisfaction and productivity (Gagné et al., 2015). The third aspect of the EI's methodology pays tribute to the EI's transcendence of the dichotomy between quantitative performance indicators and more qualitative insights into academic realms. The EI acknowledges that metrics "can give powerful insights into collaboration patterns, disciplinary orientation", but lets go of the belief that one method can "get it right"[6].

Fourth, the EI aims at an inquiry of academic value trajectories to inform both, *summative and formative* evaluations. While summative evaluations are those that "are carried out to ensure accountability for past work", formative evaluations are "primarily concerned with learning for improvement"[6]. While research assessment implicitly produce relevant lessons for everyone involved, the EI explicitly values that learning opportunity for individuals, teams, the academic institution and stakeholders. That way, the EI leads not only to a less wasteful way of using resources involved in an evaluation costs, but also bridges organisational development and evaluation (Parsons, 2009).

These four aspects of the evaluative inquiry's methodology generate "a mode of inquiry that is radically different from established evaluative protocols"[6]. Instead of starting from institutionalised norms of what counts as research output, the EI envisions a participatory approach of negotiating research value, purpose and which indicators to be evaluated on. The EI values the organisational culture as a phenomenon emerging from a density of variables and people involved and influencing knowledge production processes. It transcends the dichotomies between quantitative & qualitative evaluation methods, between scientific & societal impact and between summative & formative evaluations. It furthermore values purpose and appreciates complexity & ambiguity. The EI's methodology, therefore, can be understood as action research that aims at transforming the academic culture into a participative one. Academic organisations may become more flexible and creative. This could open the potential for more qualitative research output, but also enhanced perceptions of autonomy and trust between involved individuals & parties.

### 3. Methods

This study is based on 15 semi-structured interviews with international astronomers from various career stages and a web-based survey, distributed among astronomers worldwide. The sample selection and procedure is described in detail elsewhere (Heuritsch, 2021a & 2021b, respectively).

Interview questions used for this study involved questions, such as "What issues do you think need to be improved to guarantee better science?", "Do you feel that you are given the chance to question how science is performed?", "When did you have your first encounter with the way science is performed and assessed? How did that compare with your initial motivation to become a scientist?" and "When it comes to evaluation of research(ers), what would you suggest to be measured?". The web-based survey included, next to a comprehensive collection of quantitative questions, which results can be found elsewhere (Heuritsch, 2021b & 2021c), three so-called miracle questions. Miracle questions are open questions, which inspire the respondent to think outside the box. These miracle questions aimed at reimagining output formats (MQ1: "In an ideal world, how would you like to present your research (results) if it didn't have to be in form of a paper?"), reimagining evaluation criteria & procedures (MQ2: "In an ideal world, what would be the best way to assess your work performance?") and reimagining research (MQ3: "Imagine you were given 1 Billion Dollar for research - how would you do research differently?"). 2011 astronomers completed the survey fully and we received the following number of responses to the miracle questions: 1201 (MQ1), 1177 (MQ2) & 1205 (MQ3). Both, the responses to the interview- and the miracle questions, were coded in MaxQDA according to Mayring's qualitative content analysis (Mayring, 2000).

Interviewees (numbered IW1 to IW15) and survey respondents may be referred to as study participants or respondents. The quotations underlining the results were not chosen based on their representativeness, but their expressiveness. Hence, the number of represented quotations is also not proportional to the number of responses to a particular subject. The source of the quotations (interviewee or answer to a specific miracle question) are stated in square brackets. Since we performed a qualitative analysis, it doesn't make sense to state exact respondent numbers when introducing topics that study participants mentioned. We therefore only mention rounded numbers for illustration, where we find it appropriate.

With the limited time and resources available, we put our best effort to follow our own recommendations from the discussion section to make this paper a bit less linear. The more interactive version of this paper can be found [here](), which includes many more quotations and anecdotal statements in order to do justice to the rich and many responses, study participants invested their valuable time in.

4. Results

This section presents the results of our interviews and miracle questions. The order in which we present the results is according to the order the miracle questions appeared in the survey. The order within the three sub-sections is chosen according to the flow of the narrative. For each sub-section, the most important key words are highlighted in bold & italics and visualised through a mind map per miracle question. This is the PDF version of this publication – please find the mind maps as attachments to this PDF (*paperclip symbol on the left in Adobe Acrobat reader*). These mind maps entail hyper-links to participant's anecdotes and quotations relating to the respective keywords.

*4.1. MQ1 – Imagining different outputs: Re-evaluing the paper format*

➔ *find the mind map for MQ1 as attachment to this PDF*

The first miracle question investigates what output alternatives to a paper an astronomer would like to engage in. More than 300 study participants see ***no alternative*** to writing results up in the form of a paper, for mainly two reasons: First, many find it difficult to think of an alternative and second, many express it is simply the best way to distilling and disseminating their research. However, the sheer amount of alternative suggestions demonstrate that there is a desire for *diversity in output formats*. More specifically, study participants advocate for the freedom to ***adapt their output format to the present context***. This sub-section presents a variety of suggested alternative forms of presenting one's research than the current publication process.

First, study participants emphasise that the ***dissemination of knowledge*** is important to them. This includes reaching the community, but also outreach. Many therefore suggest a variety of easily digestible formats that shall be shared online and open access. Examples are online material/ lectures, online short articles, press releases, theatre style talks (e.g. TED talk), media appearances (news articles, TV, radio) and comic illustrations.

Second, many study participants advocate for ***improving the publication process***. They criticise that non-English papers are not received by the astronomy community, putting those scientists at a disadvantage who cannot write proficient English papers and cannot afford appropriate proof-reading. Many address at the dysfunctional peer review system, where researchers often don't have enough time next to their research to conduct it properly. Personal biases may also play a role since despite alleged anonymity of the author, quite often the peer reviewer can guess the paper's origin. Some therefore advocate for open peer review to have transparency in the process. Furthermore, participants criticise that they have to do a bulk part of the editing workload despite the journal owning the rights to the paper. Some therefore advocate for the community owning the journal instead, so that researchers can set the rules of the game themselves.

Third, study participants advocate for a ***greater focus on quality than quantity***. On the one hand this can be achieved by publishing less and potentially larger, more comprehensible papers. This would reduce the current information overload, where it is impossible to stay on top of the research and produce high quality research [IW9]. On the other hand, study participants also suggest to publish micro-publications (of partial results), which would make the dissemination of knowledge more agile. Micro-publications, however, entail the potential to increase information overload. Therefore, there is an ambivalent potential to increase research quality in both, micro-publications and macro-publications. Micro-publications have the potential to reach the community quicker and being more digestible, while macro-publications may decrease information overload and give a more complete overview of the entire path of the research process/ treatise, including failures along the way.

Fourth, ***negative results***, as well as ***data and reduction code*** shall be publishable in some form and count as output. Data and code shall be stored in some sort of repository that can be linked with each other and the results. Software and data shall become more discoverable and citable and given equal value as papers in terms of output. Right now, the development of reduction code only counts towards one's achievement if a whole paper can be dedicated to it. Furthermore, making data and pipeline code transparent would improve replicability and further advancement of knowledge, increasing research quality [e.g. IW4, IW5, IW6, IW7].

Fifth, related to a faster dissemination of knowledge and transparency, study participants value ***preprint repositories, such as ArXiv***, as a valuable source of knowledge. They suggest to make ArXiv more interactive, where peers can comment and review the presented research. Furthermore, any documentation that reflects the research process, such as ***(lab) reports, manuals, research notes and notebooks***, is perceived as a worthy output. Shared code notebooks, such as Jupyter/ IPython, shall be executable, interactive and linked to data.

Sixth, **open discussions** with peers and the community are not only valued as vital for the research process, but shall also be perceived as a form of output. This includes collaborations, seminars, meetings (e.g. with the research group), conferences, colloquia, giving talks, visits to institutions, giving workshops and short presentations. Knowledge shared in those discussions, shall be disseminated and freely shared online, such as in the form of recordings and conference proceedings.

Seventh, while some study participants point out that it's important that an astronomer is involved in the whole research process, including writing up the results, some state that it's not everyone's talent to do so in a coherent and readable way. They wish for others (**"ghostwriters"**) who can communicate results better to write the paper. Language editors and graphic designers, who polish the papers, could assist this process.

Last, but not least, many study participants advocate for **updating the antiquated paper style**, which does not live up to the technical possibilities of the 21$^{st}$ century. The current linear style is perceived as boring to read. Moreover, in order to maintain readability, a linear paper may not contain too many details, such as derivations in calculations and theoretical background. This makes it difficult to follow the reasoning; astronomers in a different subfield may not at all understand the presented research, and astronomers in the same subfield may not be able to replicate exactly what was done. IW5, a former astronomer, now working for the journal of the American Astronomical Society, argues for "transitive referencing". This is a citation system, which detects the (data) sources cited studies are built upon in order for those to also receive credit. The debate about style is closely related to the debate about whether a paper is the only way to communicate results. Many study participants advocate for a mixed style in order to present results more accessibly to colleagues and the public. Participants wish to enhance their publications not only with links to data, code and calculations used for the study, but also with visualisations and videos. Many media formats are suggested to enrich or (partly) substitute the paper: videos, books, podcasts, blog posts, social media and visualisations (e.g. plots, figures, charts, graphs, images, flow chart, VR experiences, art installations). Next to publications in multimedia formats, many study participants wish for the publication to be a "live format" ("live manner", "living papers", "dynamic" [MQ1]). This means that publications are updatable and include interaction and exchange with peers. A way to integrate multimedia and live formats is a knowledge repository in the shape of a Wiki. Study participants imagine this "encyclopaedia" to be continuously updated (including version control), hyperlinked, interactive, community-based reviewed, and available to everybody. This would include links memos, lab reports and other documentation as well as to data, code and calculations.

To summarise this sub-section, study participants find it important that knowledge gets "preserved" [IW10] in order to advance knowledge and so that nobody reinvents the wheel. While many participants think knowledge preservation in form of a paper is the only way, many others argue that they quite of lack in "digestibility". Organising knowledge in a more innovative and interactive way than the current paper style would pay tribute to many alternative output formats mentioned before, such as including open discussions and increasing replicability. Moreover, formats like Wikis would give a better overview of the current state of knowledge, reducing information overload. This includes not having to paraphrase one's own introduction time and time again when another paper is written based on the same study or theoretical background.

*4.2. MQ2 – Re-evaluing evaluation*

➜ *find the mind map for MQ2 as attachment to this PDF*

The second miracle question investigates what study participants wish with respect to what should count towards their performance assessment. This is related to the first miracle question insofar that output (currently in the form of peer-reviewed papers) is an important performance criterion. While the first miracle question resulted in a variety of different output formats that shall be recognised as part of an astronomer's performance, this question goes beyond output and expands to other aspects of research.

<u>Critique of indicators</u>

Study participants recognise the fact that quantitative performance indicators, such as publication and citation rate are perceived as more objective among decision makers than any qualitative attempt to make achievement visible, such as recommendation letters. However, many voice the concern that "in our urge to make it as impersonal or objective as possible, we have removed, uh, the personal aspect from it all" [IW1]. Many study participants share the opinion to not be "against peer-reviewed papers, [but] it is the single focus on publications and the hyper competition that is the problem. Reducing an innovative 3D human scientist to a single metric number is not a scientific way to measure a scientist!" [MQ1]. Many worry that "a system where we need to worry constantly about 'performance' is one which destroys creativity" and advocate for "less assessment of our work performance, and more freedom to develop new ideas" [MQ2].

<u>Research is difficult to be measured</u>

Some study participants go so far as to claim that research performance cannot be evaluated meaningfully, "because it is priceless" [MQ2]. This reasoning is based on the argument that fundamental research often brings delayed returns to society; "[Research] is an investment toward the future. Therefore, useful results – if any – will be available on time scales much longer than any assessment of work performance" [MQ2]. As such "it should be continuously made clear to society that science (and especially astronomy) is a long-term investment" [MQ3]. Others raise the argument that "research is not linear and cannot be [monitored] without paying the high cost" [MQ2], such as shortcuts, including questionable research practices. Many factors, such as luck, personal circumstances and support of the supervisor, are mentioned to make or break a scientific career, independently of one's scientific skills. Therefore, any assessment shall consider a variety of contextual factors and give "room for risk and/or longer reflections" [MQ2].

<u>All Aspects of research shall count</u>

Despite this critique on (quantitative) assessment of research, more than 30 survey participants explicitly state that they want to be evaluated on basis of ***quantitative indicators***, such as publications (as output to disseminate knowledge) and received grants. However, more than 50 participants advocate for a mix of quantitative and qualitative assessment and more than 40 explicitly state that they want all aspects of research included in a form of "holistic" [MQ2] way of evaluation. To "account for the complexity of the research" [MQ2] many aspects of research that do not currently count towards a scientist's record of achievements or as output shall inform a more balanced, comprehensive and bird's-eye view. The previous sub-section already described alternative output formats which may count towards an astronomer's achievements and this sub-section digs deeper into all the ***background work and intangible contributions*** which are invaluable in doing research and shall therefore be accounted for.

First, as with output, study participants find it important to ***consider context*** when evaluating someone's performance. Instead of counting one's publications, accomplishments shall be measures relative to agreed goals and dependent on the effort and commitment the researcher is demonstrating. This includes considering the astronomer's efficacy and independence in their research endeavour in the light of personal circumstances (having a family, illness, etc) and the work environment (such as its available resources).

Second, a significant type of background work is mentioned to ***be managerial-/ organisational-/ and admin tasks***. This includes community services, such as peer review and serving on telescope time allocation or grant allocation committees. Furthermore, time invested in institute activities and proposal writing – regardless of the outcome – shall be accounted for.

Third, the ***research process*** itself includes a variety of tasks, which study participants wish to count towards their performance, irrespective of the resulting paper. This includes data gathering and publication, considering the value of the collected data, the development and publication of tools, software & reduction code (as also mentioned in the first miracle question), technical contributions, developing & maintaining instruments, a detailed description of one's analysis & methodology, and appropriate visualisation of results. Just as described in the first miracle question, study participants value negative results and discussing science with peers in (online) conferences, talks, colloquia and interdisciplinary exchanges. Moreover, efforts to reproduce published research shall be acknowledged.

Fourth, study participants advocate for valuing a researcher's ***problem solving capacity*** and ***creativity***. Innovative & original ideas, lateral approaches, interdisciplinary efforts and curiosity shall be rewarded. Not going for the "low hanging fruits" [MQ2] shall be acknowledged, even if taking risks does not result into a paper.

Fifth, ***research quality***, which quite regularly is mentioned in contrast to quantity (cf. first miracle question), shall be valued in its own regard. Qualitative aspects that study participants wish to be acknowledged include the researcher's integrity & scientific rigor, such as the logical validity, completeness, consistency, precision, accuracy and scepticism with respect to performed research. Furthermore, research (results) shall be outlined in a clear and reproducible way. For example, IW1 refers to replicability as tracking *provenience* – one needs to know exactly where data, code and calculations come from and how they were used. Since astronomy is a basic science, which is little commercialised compared to others, IW1 emphasises that this field is in a unique position to allow for complete provenience transparency.

Sixth, ***knowledge contribution and impact on the field and society*** are not only an important driver to publish, but shall also be accounted for in diverse ways of contributions. The relevance, importance and usability of one's research shall be acknowledged. As mentioned as alternative output candidates in the first miracle question, study participants emphasise the value of ***teaching*** and ***outreach***. IW2 and IW8, for example, mention that the impact they can have by training students in terms of transmitting curiosity, passion and knowledge makes them proud. Study participants would like to receive recognition and support for educating the public about science and astronomy. This includes all forms of dissemination of knowledge, and science communication.

Seventh, study participants advocate for a peer assessment which goes beyond peer review. Peers include, for example, supervisors, bosses, committees, collaboration partners, experts in the field and colleagues. Since performance measurement through indicators is perceived as too superficial by many study participants, they wish for a more personal assessment, despite the potential biases that may come with it. Rethinking recommendation letters to make them more personal, and personal interviews to gauge the astronomer's capabilities of creative thinking and ambitions, are examples of what study participants recommend [IW12 & IW13].

Last, but not least, respondents value efforts towards a "healthy" **work culture** [MQ2]. This includes contributions to a more inclusive & diverse environment and team efforts, such as supporting each other and being a "team player". Collaboration & social skills and good supervision & mentoring shall be acknowledged.

To conclude this sub-section, while some study participants wish to be solely evaluated by quantitative indicators, such as the publication rate, the majority mentioned a variety of work tasks, which are not currently acknowledged as performance. Evidently, "all the things that one associates value with" [IW14], which are related to and facilitate research activities shall be recognised.

*4.3. MQ3 – Transformation: Re-evaluing values*

➔ *find the mind map for MQ3 as attachment to this PDF*

The third miracle question investigates what astronomers would choose to spend money on with respect to performing research. We worked from the assumption that study participants would choose to spend money on aspects about research they find important and/or are currently undervalued. We do find that this miracle question is related to both former ones: What astronomers would like to count (as output) is also what they value about research.

A few study participants responded that they would ***reject the money*** or redirect it, since managing that kind of amount distracts them from research. Only three participants stated that they would abandon research, when given a billion Dollars. The following sub-section outlines the various research aspects that study participants would invest that amount in.

First, many study participants would invest money into ***resources***. This includes daily work equipment (such as computing resources & laboratory equipment), supercomputers and – perhaps emphasised in COVID-19 times – home office equipment and household help. Furthermore, participants would invest money in better infrastructure, such as more efficient, centralised and curated data storage. This relates to the desire to invest into the data analysis of existing data bases and cleaning up data. Moreover, algorithms and code shall be written in a more standardised way and be accessible for easy use. About 150 study participants would invest into the construction of instruments, such as telescopes and satellites and more than 40 participants would ensure more telescope time. Many respondents would hire more researchers and support astronomers to delegate tasks related to data science, statistics, technical support, proofreading, managerial support and admin tasks. Just under 40 respondents would hire software engineers to write code, clean up existing code and take care of its maintenance. More than 80 study participants would spend money on support staff/ scientists, such as technicians or science communicators, but also behavioural scientists who reflect on the way research is done and discuss ways to improve with the astronomers. More than 130 respondents advocate for more early career researchers (ECRs), while some others argue that there are too many compared to the permanent positions available. This discrepancy may result from the wish to get a lot of research done (through cheap labour), while at the same time realising that this is not sustainable. Related to the cheap labour aspect of ECRs, some participants argue for higher/ fairer salaries, especially for PhDs and postdocs.

Second, many study participants address the poor job prospects of an ECR. Given 1 Billion Dollars, more than 140 respondents would *improve job security and invest into longer contracts* and more permanent faculty positions. IW7 mentions the job uncertainty in academia as a bigger problem than publication pressure and IW11 emphasises that the lack of faculty positions as compared to the number of ECRs is an essential issue. Furthermore, study participants would decrease the need for astronomers to relocate every few years for another position. The necessity to relocate is perceived as a stressful factor for many, especially if a family is involved. Moreover, respondents point out that moving every couple of years decreases researcher's productivity, given the organisational burden that comes with relocation.

Third, *mentoring students and ECRs* was already mentioned in the previous miracle question as an activity that shall count towards an astronomer's performance and is also worth spending a significant amount of money on. This includes helping astronomers in junior positions grow professionally, but also *providing comprehensive training* to them. Some study participants advocate for less privileged students to be especially supported by, for example, receiving scholarships. Furthermore, study participants would invest money into developing programming and social skills, as well as management and leadership training. Participants would also invest into more training in meta-topics, such as the workings of the science system and possible career paths after a junior research position. This includes teaching ECRs that it is not a failure to get negative peer reviews and to leave academia with a valuable degree [e.g. IW10, IW15]. Instead, they shall be provided with "graceful exit paths" [IW1]. The previous two miracle questions showed that teaching and outreach are activities that study participants hold in high regards and this miracle questions emphasises this, since many would spend money on the *education* of students, ECRs and faculty, as well as the public, such as through outreach.

Fourth, study participants would invest money into improving the *peer review process and funding allocation*. It was already mentioned in both previous miracle questions that performing peer review shall count as a form of output and towards the performance of an astronomer. Participants would allocate funding for paying high standard peer review and some suggest a funding lottery to make grant allocation fairer for those who propose high quality research projects.

Fifth, more than 120 respondents would invest money into *collaboration*, a research aspect which was also already mentioned to be important in both previous miracle questions. This includes travelling to foreign institutions and conferences, as well as inviting collaboration partners from abroad and investing in innovative formats for scientific exchange (across institutions).

Sixth, many study participants would use the 1 Billion Dollars to feel more *freedom in their research*. This includes feeling less publication and funding pressure, "buying [oneself] out of teaching" [MQ3] and circumventing political & bureaucratic processes. Some participants would also work less to spend more time with their family and some would spend funds on long-term, "blue-sky" research, studying big research questions. Around 150 respondents mentioned that financial freedom would enable them to focus better on (good quality) research, such as ensuring reproducibility and researching unpopular, less-funded ideas. Relatedly, respondents would feel freer to think creatively (an ability that was also mentioned to be important in the second miracle question) and would invest into funding innovative ideas (which was mentioned to be important in both previous miracle questions). This could include funding a certain fraction of risky topics [IW11].

Seventh, many study participants would invest the large sum into ***building their own institution***. This would be an institution where ***interdisciplinary research*** (an aspect mentioned to be important also in the second miracle question) & collaborations would be fostered and the above mentioned freedom to perform high quality research would be provided. Furthermore, respondents advocate for such an institute to encourage a healthy research culture, which leads to the last point.

Last, but not least, many study participants advocate for a ***culture change in academia***. They would like to have money and efforts allocated towards supporting a healthy work-life balance, taking care of researchers' mental health and creating a friendly environment for parents. Some respondents argue for more research retreats and soft skills trainings to foster a more collaborative environment. Around 100 participants would like to have more teamwork and would support this by, for example, having communal workspaces and fostering more trust in each other. Many participants would prefer an agile & flat hierarchy management and a "non-toxic" work culture [MQ3]. Related to what was found in the second miracle question, study participants would like to be evaluated differently than through performance indicators and would want to receive recognition, appreciation & acknowledgement for all aspects of their work. Some respondents would invest funds into environmental sustainability, including the advancement of light pollution mitigation. Finally, more than 110 respondents advocate for more diversity in a multidimensional sense. Instead of only focussing on diversity in gender and ethnicity, true diversity means also valuing "diversity of thought and background" [MQ3]. Furthermore, excellence and success shall also be recognised as having different dimensions than just the ones that can be measured my performance indicators [e.g. IW13]. Respondents would invest money into efforts towards supporting researchers with families, researchers from less privileged countries, efforts to reduce biases and encouraging different working styles. A diverse environment is believed to be one that doesn't only provide more fairness, but also one which encourages innovative ideas.

In conclusion, we find many ideas mentioned in alternative output suggestions (MQ1) and alternative solutions to what shall count in research in astronomy (MQ2) also reflected in the last miracle question of what astronomers would spend a significant amount of money on (MQ3). Naturally, many of the mentioned aspects are more intangible than what is currently measured by performance indicators, but the fact that so many aspects overlap shows that there is a need for acknowledgement of those aspects as well. Research is much more than the resulting papers and the richness of the respondent's answers provides us with a good basis for re-imaging the (publication) culture in astronomy.

## 5. Discussion

What stands out from the answers to the three miracle questions about imagining alternative outputs, evaluation criteria and which aspects of research are important to be invested in, is that a variety of many different aspects of research were mentioned and overlapped across the three questions. Our results show that there is a need for adapting output and evaluation to the context and that research performance shall be valued in all its aspects. Hence, a scientific culture which encourages high quality research may not only be based on intellectual & demographic diversity, but also value a diverse set and combination of different facets of the research process.

Drawing on the variety of ideas that were mentioned as a response and on solutions that are being implemented and discussed around the world in order to improve aspects of the scientific system, we arrive at the following recommendations[23] about what a desired culture in astrophysics could entail. The order is chosen according to the flow of the narrative.

*5.1. Recommendations – Towards a democratic academic culture*

5.1.1. Changing the Publication Paradigm: Replacing Academic Journals

Study participants argued that they would like to own the journal they publish in as a community to dictate the rules of the publication process themselves. This includes improving infrastructure that allows for curated data & code storage. While "replacing traditional journals with a more modern solution is not a new idea", "the lack of progress since the first calls more than 15 years ago has now convinced an increasing number of experts that a disruptive break is now necessary" (Brems et al., 2021: 1). The reason for this inaction is that every player is locked-in: the researcher, the library and the institution. Whoever moves first would be at a disadvantage and "the corporate publishers are the only player profiting from this system" (ibid.: 3). The authors estimate that 700 tools (ibid.: 7), developed by start-ups and community initiatives, already exist, which could replace traditional journals with an open scholarly standard.

The disruptive break from the lock-in needs to be one with regards to governance to strengthen the independence of science. Under the governance of the scholarly community, a modern digital infrastructure could evolve to a "decentralized, resilient, evolvable network that is interconnected by open standards, that allow seamlessly moving from one provider to another", preventing another vendor lock-in (ibid.: 5). Such open standards already exist (for examples refer to Brems et al., 2021: 6) and could be expanded on basis of continuous feedback of the scientific community. A replacement solution will not only encompass literature in paper format, as we know it, but also "all components of the scholarly workflow, with modern technologies taking care of text, data, and code, allowing dynamic updating, version/quality control, and tracking of contributor-ship" (ibid.: 6). This will require specialised staff dedicated to the curation of the new output platform and realigning financial incentives with academic interest. The latter may include updating funding guidelines to reward efforts that support and feed such an open infrastructure with functionalities and knowledge (Breuer & Trilcke, 2021; Brems et al., 2021).

Establishing such a knowledge management infrastructure will require all stakeholders to cooperate. The development of data sharing standards and solutions for technical and ethical issues, such as the sharing of confidential personal information, must be discussed and implemented by researchers, institutions and funding agencies. Miyakawa (2020) and Breuer

---

[23] Additionally, we refer to initiatives such as summits on participatory research and resolving the researcher's structural dependence on temporary contracts and to a brief overview of potential solutions to improve research culture (p.49).
Further references to scientific literature:
Anderson, M.S., Horn, A.S., Risbey, K.R., Ronning, E.A., De Vries, R., Martinson, B.C. (2007),"What do mentoring and training in the responsible conduct of research have to do with scientists' misbehavior? Findings from a national survey of NIH-funded scientists", *Academic Medicine* 82(9): p.853-860.
Gunsalus, C.K., Martinson, B.C., Faulkner, L.R., McNutt, M.K., Nerem, R.M. (2019) "Overdue: a US advisory board for research integrity", *Nature* 566: p.173-175.
Martinson, B.C., Thrush, C.R., Gunsalus, C.K. (2017, "Comment on 'Improving research misconduct policies' by Redman & Caplan", *EMBO Reports* e201744295

& Trilcke (2021) summarise what a non-commercial, community-governed, open infrastructure would require (last two points added by the author):

- expanding suitable publication platforms,
- developing standards for the publication of new formats,
- organising quality assurance processes,
- modifying remuneration mechanisms,
- clarifying legal questions,
- training and recruit specialised staff,
- hiring software developers, specialised in machine learning, and
- inquiring continuous feedback from the scientific community to make sure that functionalities are based on their needs to present output accurately.

Our study is congruent with findings of Zuiderwijk & Spiers (2019) that astronomers feel motivated to share data & code for the sake of transparency and speeding up the research process. Open source coding generally has a decades-long tradition in academia (Brems et al., 2021: 7). Zuiderwijk & Spiers (2019) find that demotivational factors for open sharing of information are significantly related to a lack of (interlinked) infrastructure. Providing such an open infrastructure in the form of a "global library of interoperable repositories" (Brems et al., 2021: 7), not only moves away the focus from the "paper" as the only output that counts, but also enables researchers to share other valued aspects of the research workflow, such as data & code. Evaluation of a researcher's performance may be adapted as a consequence and scientists may be able to focus on quality, rather than "wasting time" on antiquated publication processes. Therefore, bringing the "means of scholarly communication back under the sovereignty of the scholarly community" (ibid: 6) by replacing the journal system with a digitally innovative solution, would change the publication paradigm. This paradigm change could contribute towards a solution of several aspects addressed by our study participants: the requirement for various output formats and for data & code repositories, the need to improve peer review, the quality of published research (especially in terms of replicability) and the need for a diversity in evaluation criteria. The following recommendations will elaborate on these and more aspects for improvement in academic culture in astronomy, raised in this study.

<u>5.1.2. The Living Knowledge Repository: Adapting output to context</u>

The sub-section above argues for replacing the traditional academic journal by a community-governed, open infrastructure, which enables the researcher to communicate all stages of the research workflow. This sub-section focusses on what such a living knowledge repository may entail.

Despite the fact that "communication in science is an integral part of the process of 'doing science'" (Albrecht, 2007: 141), publications are increasingly becoming more difficult to understand (Hayes, 1992), and are based on antiquated "Gutenberg technology"[24] (Albrecht, 2007: 143). An open scholarly infrastructure would offer a living knowledge repository instead. Other labels found in the literature are "3D publication model" (ibid.: 143), "Wikipedia-type of representation" (ibid.: 148), enhanced publication, and interactive, integrative, dynamic, hybrid formats including multimedia (Breuer & Trilcke, 2021). Whatever the label, such an open infrastructure would come with several advantages:

---

[24] In fact, it is quite astonishing, that scientists were the drivers behind the advent of the internet, starting with the ARPANET in the late 1960s, while now scientific publications lag behind technological possibilities. The refereeing- and the output system are a few hundred years old (Moosa, 2018: 137).

First, this new platform would offer functionalities that a linear print format cannot, such as access to original data, to literature references, transitive & forward references, links to supporting information, calculations and so forth. Depending on the context and what makes sense in terms of efficient communication, researchers could exploit the diversity of available media formats, such as animations, videos, podcasts and collapsible text.

Second, a Wikipedia-type of representation bears the potential to counteract information overload. Because information can be easily interlinked, one does not have to increase information redundancy and waste time by paraphrasing parts of (one's own) previous papers in order to avoid (self-)plagiarism.

Third, a living knowledge repository would allow for dynamic updates. Researchers could present their latest stage of research and possibly even what-if-scenarios/ hypotheses (Albrecht, 2007) and update information once it has become obsolete. Those stages of research which "used to be pre-publication" (Breuer & Trilcke, 2021: 5) could be presented in dynamic formats, reflecting on "the fact that research is in itself a process" (Breuer & Trilcke, 2021: 8). A dynamic format is never final, but is a numbered version that can be updated, archived and exchanged.

> "Dynamic updating and version control with persistent identifiers allow scholarly outputs to move from a static 'version of record' to living outputs which can be rapidly updated to reflect the best available knowledge at any time." (Brems et al., 2021: 8)

Fourth, the 3D publication model would change the "chronological order of, and linkages between, individual stages of research" (Breuer & Trilcke, 2021: 5). Sharing data and research stages that used to be pre-publication could result into "network-shaped bifurcations between the output of these former pre-publication stages and the traditional formats used for the publication of results" (ibid.: 5). Particular results may become obsolete, while a publication of an earlier stage of the research remains valid.

Fifth, neural network and machine learning techniques could assist humans in full text analyses to identify relations amongst research and overcome language barriers. Instead of a traditional literature review, an AI solution could combine information from different sources, in order to present the latest knowledge on a specific subject in an encyclopaedia style (Albrecht, 2007).

Sixth, sharing various aspects of research (stages) increases visibility of tasks and scientific information that was formerly not available. This may not only increase transparency and replicability, and consequently the quality of research (cf. *section 5.1.5.*), but also create a multitude of achievements to be evaluated on (cf. *section 5.1.7.*). Furthermore, this increases the potential to make science more accessible and the speed of communication to the scientific community, but also the public (Breuer & Trilcke, 2021).

### 5.1.3. Diversifying Peer Review: Open & Community Review

Our study's findings agree with conclusions of other literature that the current peer review system in science is flawed and perhaps only "provides an 'illusion' of 'quality control'" (Moosa, 2018: 137). The process from submission to publication in a conventional journal may take several months to a couple of years (Albrecht, 2007; Moosa, 2018). This impedes knowledge dissemination, which is further decreased by paywalls imposed by the commercial journals (Moosa, 2018).

> "We have to remember that under the current system, peer-reviewed journals operate against the principle that research is intended to benefit the society." (Moosa, 2018: 179)

Replacing the journal system by an open, community-governed infrastructure, would answer to the many calls for replacing the peer review system with an open repository system, where communication of research (results) is accompanied by a review and evaluation[25] of output (e.g. Moosa, 2018). Such an infrastructure would allow direct author-reviewer interactions, where anybody in the community who can make valuable remarks or contributions could assume the role of a reviewer. The paradigm may shift away from pre-publication peer review to post-publication peer review of any kind of output on any kind of research stage. A review then may be a more complex interaction than it is now; consisting of discussions, commentaries and endorsements (Moosa, 2018). Transparent and AI-assisted review would be as possible as (pseudo-) anonymous peer review, depending on the needs of appropriate quality control (Brems et al., 2021: 8).

Community peer review within an open infrastructure may entail the following advantages (cf. Moosa, 2018): (1) The collective wisdom of the scientific community is harnessed. (2) The evaluation process is accelerated and expanded with new forms of credit (Munafò et al., 2017). (3) Speed and access of information flow increases. (4) Information redundancy and (time) resources are saved by information being centralised and not distributed among many different journals. (5) The relevance of the research is determined by the interest and responses it generates, rather than by journal impact factors or secretive editor/ peer reviewer decisions.

> "We should think about our field like a marketplace of ideas. Everyone should be free to put their ideas out there. There is no need for referees. Good ideas will get recognised, used and cited. Bad ideas will be ignored." (Moosa, 2018: 135)

### 5.1.4. Creating an error-culture: Rewarding risky ideas and the publication of negative results

This and previous studies by Heuritsch (2019 & 2021a-c), and commentaries by renown astrophysicist Loeb (e.g. 2010, 2013 & 2014), emphasise the importance for research in astronomy to engage in studies that may not produce "publishable results" for conventional publications. Examples are studies that employ "risky ideas", which have a considerate potential to fail or to produce negative results. Since research is the endeavour to study the unknown, however, it is important that failed paths and negative results are shared with the community. Moreover, research[26] (e.g. van Dyck et al., 2005; Frese & Keith, 2015) has shown that a positive error-culture likely fosters creativity, because it encourages calculated risk-taking, out-of-the-box thinking and involves intuition. In this kind of culture, employees and colleagues are not punished for making mistakes, but encouraged to learn from them. Next to fostering such a culture, there are several examples for ideas and initiatives about what to do to encourage risk taking. They address mainly three areas: funding, telescope time allocation, and pre-registration of research:

---

[25] See additionally: https://www.tagesspiegel.de/wissen/forschende-begutachten-forschende-digitalisierung-schafft-neue-optionen-fuer-wissenschaftliche-qualitaetssicherung/26631890.html

[26] See also exemplary anecdotes from scientists:
TED talk by astronomer and science communicator Phil Plait (https://tedxboulder.com/videos/failing-upwards-science-learns-by-making-mistakes)
Albert Szent-Gyorgyi in *Perspectives in Biology and Medicine* (1974) 18, 41-43: "Scientific research is, in many ways, related to art.", "Projects are nonsense. I don't think that any of the great discoveries were even made by projects. They were made by intuition." (https://www.queensu.ca/academia/forsdyke/peerrev0.htm)

First, with regards to funding, a different allocation could decouple job security from productivity. This could be done by funding individuals instead of specific projects (Moosa, 2018). Another possibility is to innovate the way that productivity is measured itself, including rewarding risk tasking, as suggested by the results of our study. Other initiatives involve funding innovative ideas[27] or funding lotteries[28] that may make for a fairer distribution of grants towards equally good proposals.

Second, in analogy to funding, there has been many suggestions to allocate telescope time to a certain fraction of risky proposals (e.g. Patat et al., 2007; Loeb, 2014). This could include optimising observation time schedules, increasing data delivery performance and optimising resource distribution (Patat et al., 2007). These could be enabled by a more efficient knowledge management system, such as the open infrastructure described above (*section 5.1.1.*).

Third, pre-registering research may give exposure and space to publish research that is not currently publishable. Some funding initiatives already require pre-registration (see for examples: Munafò et al., 2017: 4). "Results-free" review, "where editorial decisions to accept are based solely on review of the rationale and study methods alone" (Munafò et al., 2017: 6) could be another way of encouraging the publication of negative results.

5.1.5. Reflexive Quality: Understanding and fostering research quality

"Recommendations in relation to quality are difficult to make. The opaqueness and positivity of the concept make misunderstandings likely." (Dahler-Larsen, 2019: 218)

Arguably, elusive concepts such as happiness and quality cannot be targeted directly without unintended consequences. Rather they are by-products of choosing values, by which one lives or works, well and wisely. Instead of measuring research quality by proxies, such as the citation rate, it may be more beneficial for science, to 1) understand the concept of quality from a meta-perspective and 2) to facilitate a culture and an infrastructure which fosters research quality in a variety of its intangible aspects.

Dahler-Larsen (2019) refers to several aspects which come with understanding quality from a meta-perspective. First, as pointed out in *section 2.1.*, quality is a reflexive concept, which means that by formulating quality inscriptions, the nature of the underlying reality is reduced to these criteria. This has constitutive effects on how reality is perceived and what actors orient their behaviour towards.

Second, the author encourages everyone to acknowledge their "part in the production of constitutive effects and [to] act accordingly" (ibid.: 219). While one may not have control over the generation of a quality inscription, one should not deny responsibility for participating in "the micro-politics of qualitization" (ibid.: 219). This also involves to not use the term "unintended effects" as an excuse: To act responsibly is to be curious about the constitutive effects of qualitization and the "ongoing systemic effects of your own actions" (ibid.: 219).

Third, Dahler-Larsen (2019) advises to "study qualitization in practice and report the findings" (ibid.: 221). Next to considering the resources qualitization costs, qualitization processes may be studied by employing an evaluative inquiry. A participative dialogue about feedback-loops would not only foster a deliberative democratic environment, but also would draw on the reflexivity of the qualitization: "If quality configurations became more self-reflexive, perhaps more intelligent and sensitive forms of qualitization would be developed" (ibid.: 226).

---

[27] For example: https://www.nsf.gov/bfa/dias/policy/merit_review/mrfaqs.jsp#10
[28] For example: https://scilog.fwf.ac.at/en/article/10632/support-risk-takers

Fourth, the author advocates for more diversity in qualitization. The availability of a rich set of quality notions does not negate the fundamental fact that quality cannot be directly measured, but underscores the relativity, ambiguity and context-dependence that comes with a complex (social) world. The acceptance that many perspectives may be equally valid and that "there is no such thing as total quality" may enhance deliberation, reflection (ibid.: 79) and ambiguity tolerance.

Lastly, Dahler-Larsen (2019) argues for considering alternatives to qualitization. In an ongoing deliberative process, quality inscriptions may come and go, adapt and are applied in specific contexts. Moreover, a participative dialogue counteracts two shortcomings of qualitization: Qualitization "buries problems" and it "does not recognise the reasonableness of its opposition" (ibid.: 228). By talking about organisational or work-related problems instead of negating them, as they are not reflected by quality inscriptions, more helpful solutions may be found. Likewise, by having controversies about value conflicts, the involved parties may find some value in the aspects that are not measured by the currently valid concept of quality.

Fostering an error-culture and providing an open knowledge management infrastructure, as described in sub-sections above, has the potential to mitigate two crises related to research quality, which were identified by Brems et al. (2021) and also addressed by our study participants: the replication crisis and the functionality crisis. The rest of this sub-section is dedicated to elaborate on this potential.

The lack of reproducibility in science is partly caused by the lack of the infrastructure and incentives to publish raw data, code, negative results and other aspects of the research process that do not fit the narrative of the resulting paper. If data could be shared in a systematic manner within an open infrastructure, reanalysis and data mining would be made possible and simplified (Miyakawa, 2020). The author does not mince words when he calls for a "no raw data, no science" policy. Equally, incentivising the use of this infrastructure for optimal sharing is an important step in the shift from an individualistic to a collaborative scientific culture. Studies by Heuritsch (2019 & 2021a-c) and Zuiderwijk & Spiers (2019) show that astronomers generally like to share their data out of autonomous motivation, but often refrain from doing so because it takes valuable time away from improving their publication count. Hence, we conclude that encouraging sharing behaviour by making it count towards one's performance may already be the sparking nudge[29] necessary.

Additionally, antiquated publication processes and the dependence on a journal for publication, have caused a "functionality crisis" (Brems et al., 2021). Digital tools, which either already exist or would technologically be possible to develop, are lacking their application. This not only means that researchers may find themselves wasting a lot of their time on procedures that could otherwise be done much quicker, but that the replication crisis is fed by comparatively slow processes and the lack of interconnected and shared research (data). While prestigious journals may add to the replication crisis by "capitalizing on surprising, too-good-to-be-true results and lacking proper quality controls" (Brems et al., 2021: 3), an open infrastructure which allows for publication of all sorts of aspects of the research process would tackle this and the functionality crisis at the same time. This could be a huge step towards fostering research quality.

---

[29] Munafò et al. (2017: 7) list "promising examples of effective interventions for nudging incentives" for further reference.

5.1.6. Talent management: Responsible metrics & deliberative alternatives

From there being "no such thing as total quality" (Dahler-Larsen, 2019: 225) follows that "there is no such thing as the one 'true' indicator" (Kurtz & Henneken, 2017: 695). All metrics are context-dependent proxies, are therefore indirect, have error and are meaningless outside their specific contexts (ibid.). The authors tested the predictive capabilities of future performance of the three methods peer review, citation ranking and download ranking and found that there is large statistical scatter in all three methods, as well as systematic bias. Awareness about the fact that every method, not only human judgement, has their systematic biases and their operational contexts, could help in making a decision when and when not to apply a specific indicator (Hicks et al., 2015). For further reference, Glänzel & Wouters formulated "20 Dos and Don'ts in individual-level bibliometrics"[30] (*find the "20 Dos and Don'ts in individual-level bibliometrics" as attachment to this PDF*).

In addition, at a conference[31], Wouters presented a list of criteria for responsible metrics:
- Robustness: basing metrics on the best possible data in terms of accuracy and scope;
- Humility: recognizing that quantitative evaluation should support – but not supplant – qualitative, expert assessment;
- Transparency: keeping data collection and analytical processes open and transparent, so that those being evaluated can test and verify the results;
- Diversity: accounting for variation by field, using a variety of indicators to reflect and support a plurality of research & researcher career paths;
- Reflexivity: recognizing the potential & systemic effects of indicators and updating them in response.

Attempting to account for the complexity and relativity of research quality and performance, there has been many calls for using bibliometrics in combination with qualitative methods in research evaluation (e.g. Havemann & Birger, 2015; Hicks et al., 2015). This could entail, for example, capturing aspects of the research process which do not currently count (see *section 4*) by creating a more complex individual research portfolio and enabling deliberative dialogues about the more intangible research aspects.

An example for such a research portfolio is the ACUMEN project[32]. A researcher's ACUMEN portfolio combines multiple quantitative and qualitative aspects about their career achievements and various impacts on the community (research & teaching) and the public (outreach & spinoffs). The concept includes giving the researcher a voice in evaluation processes with regards to telling their own narrative about their past contributions and future goals. By doing so, the ACUMEN portfolio is a vehicle for the development of evidence-based arguments about what is valuable research output and about important contextual factors, such as the stage of the career, the discipline and the country of affiliation.

---

[30] Presented at a panel discussion at the ISSI conference (2013) in Vienna (Austria):
https://www.issi-society.org/publications/issi-conference-proceedings/proceedings-of-issi-2013/
https://www.researchgate.net/publication/301853186_The_dilemmas_of_performance_indicators_of_individual_researchers-An_urgent_debate_in_bibliometrics
https://de.slideshare.net/paulwouters1/issi2013-wg-pw
[31] Presentation by Paul Wouters on "Alternative ways to measure quality and relevance" at the conference "Quality and Relevance of Research" (http://researchrelevance.nl/) on 31st January 2017.
[32] http://research-acumen.eu/

Pluchino et al. (2018) demonstrate that "the most successful agents are almost never the most talented ones" (ibid.: 23), but owe their success to luck, environmental factors and the Matthew effect. The Matthew effect denotes the 'richer getting richer' mechanism, which in science stems from the "naïve meritocratic narrative" (ibid.) that successful people are also the most talented ones and hence funding and resources flow in their direction. Consequently, naïve meritocracy fails to support the potential of the most talented researchers. That is why random resource allocation, such as funding lotteries often bring about better output than that targeted at the most successful researchers. Next to random distribution, the authors tested several strategies to "counterbalance the unpredictable role of luck" (ibid.: 23), which are aimed at making resource distribution fairer and increasing the diversity of research ideas and thereby fostering innovation. Less focus on "excellence" and more on diversification of (risky) ideas would provide the less lucky but talented researchers with new chances to express their potential, while at the same time supporting serendipity. Additionally, a culture which includes a stimulating environment, flexible resource distribution and mentoring tailored to the needs of the mentees, is likely to foster the talents of the researchers involved (ibid.).

With respect to talent management, perhaps it would be more beneficial for the researchers and for science to give them more freedom in creating their own role. If individuals could do more of what they are good at (whether that is software development, analysis, paper writing, teaching, etc.) and less of what they are not, an environment is created where (human) resources likely operate more efficiently and where researchers may flourish at the same time.

### 5.1.7. Reflexive Evaluation: Participative & Iterative

The results of this study emphasise the importance of the awareness that no form of evaluation can be totalizing or definite. Concepts of quality or performance are social constructs, dependent on history and context. The only way we can make evaluation procedures supporting science in the best way possible is, if we let those who are evaluated participate in iterative steps to continuously adapt and improve evaluation processes and criteria. If we want truly effective evaluation, which aims at good quality science and fostering researcher's talents, we seem to have no other choice than making it democratic and experimental.

> "True understanding of how best to structure and incentivize science will emerge slowly and will never be finished. That is how science works. The key to fostering a robust metascience that evaluates and improves practices is that the stakeholders of science must not embrace the status quo, but instead pursue self-examination continuously for improvement and self-correction of the scientific process itself." (Munafò et al., 2017: 7).

There are emerging attempts to introduce deliberation and relativity into evaluation processes, such as the evaluative inquiry (see *section 2.4.*). By involving not only stakeholders, but also researchers, in determining what scientific value is and how to make it visible, the EI is a participative approach to research evaluation. Applied iteratively, it may be a reflexive approach as well: a follow-up EI may capitalise on the constitutive effects the previous EI sparked. This includes valuable learning opportunities for everyone involved about what works well and what can be improved in fostering a healthy research culture and good quality research. An artistic illustration of the iterative application of an EI is illustrated in *Figure 1*.

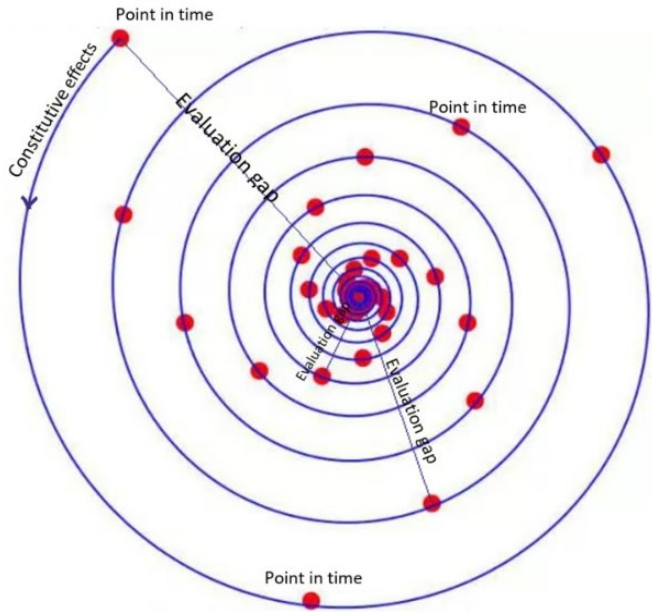

**Figure 1: The asymptotic evaluation spiral.** An evaluative inquiry is being applied "continuous in time rather than appearing in discrete moments." (Dahler-Larsen, 2019: 200). Constitutive effects of resulting evaluation procedures could be utilized such that what is valued by an evaluation procedure asymptotically approaches what astronomers value as good quality research (the centre). The divergence between those values at a given point/ period in time represents the evaluation gap. This artistic illustration was taken from Heuritsch (2018) with the author's permission.

Diversifying participation in evaluation processes may be equally important as diversifying what counts as research output and performance. An open knowledge management infrastructure may therefore be an important contribution to enable a plurality of evaluation criteria.

> "New models for tracking of contributorship and provenience will complement or even replace traditional authorship and help establish credit for valuable scientific activities not traditionally counted, such as generating and managing research data, peer reviewing, brainstorming hypotheses, or incremental improvements to open software." (Brems et al., 2021: 8)

Not only the development of this infrastructure requires all stakeholders cooperating, but it is also the responsibility of everyone involved in the research process to contribute to a healthy research (evaluation) culture.

> "Funders, publishers, societies, institutions, editors, reviewers and authors all contribute to the cultural norms that create and sustain dysfunctional incentives. Changing the incentives is therefore a problem that requires a coordinated effort by all stakeholders to alter reward structures." (Munafò et al., 2017: 7)

5.1.8. Towards a participative culture

We have learned from this study and many others what a healthy research culture, fostering content researchers and good quality research, may entail: A self-organised knowledge management infrastructure and fair evaluation procedures, which involve participation of all involved parties, leaving little room for power differentials based on academic rank and formal hierarchical structures (cf. Martinson et al., 2010 & 2013; Haven, 2021). Trust & freedom to explore one's own role and mentoring to foster potential instead of targeting performance may be as important as experiencing organisational justice and a constructive error-culture. Giving researchers the chance to learn from their (and other's) mistakes, rather than punishing them, potentially leads to more transparent science. The acceptance that our (social) world is inherently volatile, uncertain, complex and ambiguous (VUCA[33]) is a good basis for a culture supporting self-organisation instead of investing resources fighting this nature of social organisations. Reflexivity is an important part of such a culture, both in the sense of facilitating feedback loops of research processes, infrastructure and evaluation and in the sense of being aware of the constitutive effects that qualitizations and measures have. Further aspects of a healthy research culture we found to be transparency, openly discussing expectations, constructive conflict management and having sufficient time for the actual research work. These aspects are not only congruent with those found in other studies (cf. Haven, 2021), but also with a participative culture. Such a culture has the potential to address many of the wishes for change, mentioned in the miracle questions, such as the call for diversity & inclusion, trusting collaborations, good quality research & sustainability. The following paragraphs elaborate on how a participative culture may foster these aspects, followed by a paragraph on how an innovative feedback infrastructure could support the transformation from the current to a more participative culture in academia.

A participative culture has the potential to acknowledge that one-size-fits all approaches are generally not the solution to tap into people's strengths and autonomous motivations. By contrast, tailored interventions (Haven et al., 2019a) recognise and appreciate the (intellectual) diversity in human beings and groups. For example, perhaps unsurprisingly, publication pressure is more likely to be perceived by researchers ranked below full professor (Heuritsch, 2021b). Tailored interventions could entail mentors supporting early career researchers with coping with strains related to publication pressure, which are shown to be conducive to research misconduct (Martinson et al., 2010; Tijdink, 2014). Rather than just coping with publication pressure, such interventions may even lead ECRs to introduce positive changes in unhealthy aspects of the research culture (Martinson et al., 2010). Tailored interventions may also be applied to support women in science, such as improving promotion prospects and family support (cf. Fohlmeister & Helling, 2012). However, one should be aware that a simple replacement of norms or quotas is not going to lead to a more inclusive and diverse participative culture. Rather it is about transcending norms, by shifting the focus towards greater awareness, enduring ambiguity & discomfort and "fulfilment of potential for all department members" (Johnston, 2019: 1065).

A participative culture may transcend the community versus individuality dualism, by fostering teamwork, while at the same time attending the needs and interests of the individual (cf. Johnston, 2019). Aligning these interests with the overall purpose and values of the organisation is likely to be important for healthy and productive teams (Schnell, 2018). Pairing purpose with intellectual diversity (daring to think differently and questioning the status quo) and psychological safety (which is a prerequisite for the former) delivers the three ingredients for a healthy and productive team[34]. Psychological safety correlates with a good error-culture, which values growth over judgement.

A participative culture may socialise researchers into research integrity practises, such as transparency and open & faithful exchanges. Studies on the connection between the organisational culture and research integrity show that research misconduct is not so much due to "bad apples" (individual's distorted psychology; Haven & Woudenberg, 2021), but rather due to cultural factors, such as the perception of organisational injustice and the misalignment of purpose and incentives (e.g. Martinson et al. 2006 & 2010; Munafò, 2017). Research integrity, hence, "is not something someone learns from one course" (Haven et al., 2019a: 11), but it is rather a habit that comes from a culture that fosters research integrity and the psychological safety to make mistakes.

---

[33] Vision – Understanding – Clarity – Agility (VUCA') may be a handy acronym, describing a constructive approach to our VUCA world (Volatility – Uncertainty – Complexity - Ambiguity); https://www.pocketbook.co.uk/blog/2019/12/17/vuca-and-vuca-prime/

[34] See additional information by an organisational psychologist: https://ideas.ted.com/the-3-things-that-great-teams-have-in-common/

A participative culture has the potential to transform academia into an organisation that is sustainable and thereby shape a sustainable society. This potential stems from the paradigm shift from formal top-down management towards participative practices and from independent (collections of) actors towards a "systemic perspective which encompasses the complex interdependence of individual, social, cultural, economic and political activities and the biosphere" (Grecu & Ipiña: 16). Such a holistic & participative approach to labour may not only inspire democratic, but also sustainable practices within academia. By the power of example and through a spillover-effect[35] of democratic attitudes at the work place into the private life, a democratic academia could inspire sustainable & democratic life styles in society (Fricke: 225).

> "A sustainable world cannot be created without the full involvement of all members and democratic society, a sustainable world without participation and democracy is unimaginable." (Grecu & Ipiña: 15)

A transformation process from one organisational paradigm to another is a chaotic and challenging process. There may not be a blueprint for a transformation process to a participative culture, given its high complexity and context-dependence. However, this does not mean that the transformation process, escapes a realm, where scientific methods can give us evidence-based insights into what works well and what can be improved. An infrastructure, which facilitates periodic self-assessments and feedback would be an invaluable support for evidence-based transformation processes and participative cultures. A tool, initially developed by psychotherapy researchers, the Synergetic Navigation System (SNS[36]), has the potential of providing that. With this tool, tailored and periodic questionnaires could be easily distributed among all members of the organisation for a periodic assessment of all kinds of aspects regarding the working conditions. This may range from a longitudinal study about the experience of publication pressure and its influence on work, to assessing the culture around research integrity and evaluating the impact and success of customised organisational change initiatives or educational interventions. This collection of reliable and comparable feedback data to stimulate internal discussions and to inform the transformation process is recommended by several studies regarding the relationship between the organisational culture and research integrity (e.g. Martinson et al., 2010 & 2013; Haven et al. 2019b). Hence, not only can this data be used for making evidence-based decisions, but also for engaging in participative dialogue[37].

---

[35] The spillover thesis was developed by Pateman (1970). Through participation within the context of the professional life, participative-democratic attitudes & competences, as well as a positive sense of democratic self-efficacy arise. These serve the employees also beyond the work context; they grow towards being more democratically competent citizens. At the same time, this learning curve is beneficial in an environment, in which adults spend the biggest part of their lifetime. Thus, professional life is where we can learn in a microcosm what it means to participate and to co-create. By contrast, society as a whole and (international) politics is easily overwhelming without democratic training in local learning units, such as the work life at an organisation.

[36] https://www.ccsys.de/

[37] "As we use the term here, a dialogue is a collaborative inquiry through which the dialogue participants try to become wiser together, a collaborative inquiry characterized by sharing, daring and caring. In our dialogic process, to share means that we are willing to let others partake in our knowledge and vice versa. To dare means that we are willing to take risks and question our own basic assumptions and those of others. To care means that we participate with open and good intentions, each one of us as an equal among equals." (Kristiansen & Bloch-Poulsen 2005: 11)

This, according to action research, opens up the potential for everybody to learn from each other, to develop ambiguity tolerance, and finally, to democratise the organisation (Fricke, 2014: 217).

## 6. Conclusion, Implications & Outlook for Further Research

Research on research misconduct, as introduced at the outset, has shown that there is a fourth option to "love it – change it – leave it". We would call this off-the-grid option "cheat your way through the game".

Since bad apple[38] theories are too simplistic to explain deviant behaviour, such as research misconduct (from questionable research practices to outright fraud), this study made a contribution to organisational culture studies. More specifically, by employing interviews and analysing open survey questions, we qualitatively studied cultural aspects in astronomy. Setting the basis for an evaluative inquiry and future action researchers, we asked astronomers (the practitioners of this field) to reimagine output formats, evaluation procedures and research itself. While there were some astronomers who advocated for no change or claimed that the system is not perfect, but there is no better way, the sheer amount of alternative suggestions to the status-quo was overwhelming. To say it in Haven's words, we found that "the research process should be considered *in its totality*" (Haven, 2021: 177; italics in the original). Not only do the many pieces of alternative solutions give us a better image of what the research process in its totality involves, but also many participating astronomers expressed explicitly that a paradigm shift in research culture is due.

In summary, this study found that astronomers wish for a diversification and (digital) improvement of output formats, evaluation procedures and peer review. Depending on the context, output in various forms, such as blog posts, data & reduction code, giving talks and the publication of negative results is said to be valuable contributions to the community and society. Ac-*counting* for many more aspects of the research process than what performance indicators currently measure, such as alternative output formats, managerial tasks or testing out risky ideas, would make evaluation procedures value research(ers) as a whole. This is an important aspect of the culture change that many participating astronomers proposed. The need for a more inclusive and healthy research culture, including interdisciplinary and autonomous research, was expressed.

Based on the many impressions from our study participants and other studies & initiatives, we concluded with recommendations on how to shift from the current organisational paradigm towards a more participative one, which is more democratic and inclusive. A participative culture may entail trust and openness, which is what many respondents wish for. An open knowledge management infrastructure, which contains and interlinks all forms of diverse output, may be the future of science communication. Such a living repository of knowledge and data bares the potential to provide rapid dissemination of knowledge to the community and the public, improved replicability & accessibility, reducing information overload and introducing innovative ways for doing peer review. All in all, such an infrastructure would likely foster better quality research and a more collaborative culture in astronomy.

---

[38] "This is not a case of a few bad apples — it's all the apples."; Lauren Chambers about inequality issues in astronomy; in "A Breakup Letter With Astronomy, From a Young Black Woman" (https://onezero.medium.com/a-break-up-letter-with-astronomy-from-a-young-black-woman-a30de24fe209)

We propose future action research to be done on the basis of this and previous studies' findings on what astronomers value and would like to change about their research culture and organisational structures. While our sample selection did not target representativeness, the size of the sample and the variety in the responses leads to our assumption that our findings may in large part be applied to astronomy as a whole. Given the congruence with the findings of other studies, we suppose that many of our findings may even be generalizable towards the academic culture as a whole. We advocate that it is time to move beyond the meta-research performed in the field of science studies, which conventionally studies its objects and subjects from a "fly on the wall perspective" (Fricke, 2014), to a self-reflexive action research, which involves the participation of social scientists, scientists from other field and stakeholders. Action research may inspire the transformation of science towards a more democratic endeavour. Practitioners (such as researchers from a specific field) may adopt self-reflexive methodologies, such as the evaluative inquiry, in order to continuously applying its principles to move forward with this transformation on their own accord.

As Munafò et al. (2017: 7) state "the challenges to reproducible science are systemic and cultural". The involvement of many different stakeholders – including beneficiaries of the publish-or-perish culture (Moosa, 2018), such as traditional journals or even established scientists, who benefit from the Matthew effect (Pluchino et al., 2018) – makes it difficult to break out of this locked-in culture. Hence, we are not arguing that the transformation towards participative leadership is easy, nor that it would not entail any (disadvantageous) constitutive effects, which are difficult to anticipate. However, by accepting our VUCA world and that social organisations are inherently self-organised, we may find ways to appreciate uncertainty and to constructively deal with it deliberately. Participative organisations have the potential to transcend the dichotomy between autonomy and leadership that science has faced ever since it has become a profession. In order to transform, we need to be aware that the "bad apple narrative" is as simplistic as the linear machine like model of an organisation. We need to face that metrics are a form qualitization, which strips off the complexity from reality to create the illusion of certainty and objectivity. We need to embrace diversity, reflexivity, deliberative democracy, experimentation and ambiguity.

You may not need to love it, change it, leave it, nor cheat your way through the system. By simply being the change you want to see in the word, you contribute to its transformation. *Transform it!*

**Funding:** This study was performed in the framework of the junior research group "Reflexive Metrics", which is funded by the BMBF (German Bundesministerium für Bildung und Forschung; project number: 01PQ17002).

**Institutional Review Board Statement:** Not applicable.

**Informed Consent Statement:** Not applicable.

**Acknowledgments:** First: I would like to extend my gratitude to the 15 interviewees and 2011 astronomers who likely dedicated way beyond half an hour—despite the publish-or-perish imperative—to complete this survey. I am especially grateful to the astonishing fraction of >1/2, who took the time to respond to the three miracle questions quite elaboratively. Second, Natascha Kostial was a constructive sparring partner throughout the analysis and performed the inter-coding in order to increase code reliability. Third, I would like to thank my supervisor, Stephan Gauch, for facilitating this project. Fourth, thank you to Andreas Zeuch for proofreading and our discussions on how the complex and rich terms used in this paper may be introduced in a readable and understandable way without undermining their complexity and ambiguity. Last but not least, I would like to thank Julien Baart for his mental support in the phase of writing this paper and our many inspiring meta-talks.

**Conflicts of Interest:** The authors declare no conflict of interest.

**References**

Albrecht, R. (2007), "Plethoric prose vs. salient points", In: Heck, A., Houziaux, L. (eds.), "Future Professional Communication in Astronomy" (p.141-149), *Mém. Acad. Royale Belgique in 8°*

Anderson, M.S., Ronning, E.A., De Vries, R., Martinson, B.C. (2007), "The Perverse Effects of Competition on Scientists' Work and Relationships", *Sci Eng Ethics* 13. p.437-461, https://doi.org/10.1007/s11948-007-9042-5

Bleiklie I. (2018), "New Public Management or Neoliberalism, Higher Education". In: Teixeira P., Shin J. (eds.), "Encyclopedia of International Higher Education Systems and Institutions", Springer, Dordrecht. https://doi.org/10.1007/978-94-017-9553-1_143-1

Boes, A., Kämpf, T., Lühr, T., Ziegler, A. (2018), "Agilität als Chance für einen neuen Anlauf zum demokratischen Unternehmen?", *Berlin J Soziol* 28: p.181-208, https://doi.org/10.1007/s11609-018-0367-5

Bouter, L.M., Tijdink, J., Axelsen, N., Martinson, B.C., ter Riet, G. (2016), "Ranking major and minor research misbehaviors: results from a survey among participants of four World Conferences on Research Integrity", *Res Integr Peer Rev.;1(17)*: p.1-8, https://doi.org/10.1186/s41073-016-0024-5

Brems, B., Huneman, P., Schönbrodt, F., Nilsonne, G., Susi, T., Siems, R., Perakakis, P., Trachana, V., Ma, L., Rodriguez-Cuadrado, S. (2021), "Replacing academic journals", https://doi.org/10.5281/zenodo.5526635

Breuer, C., Trilcke, P. (2021), "Expanding academic publishing practices alongside the digital turn". Edited by the "Scientific practice" working group of the Priority Initiative "Digital Information" by the Alliance of Science Organisations, https://doi.org/10.48440/allianzoa.042

Coghlan, D., & Brydon-Miller, M. (2014) "The SAGE encyclopedia of action research", (Vols. 1-2). London: *SAGE Publications* Ltd, https://doi.org/10.4135/9781446294406

Crain, L.A., Martinson, B.C., Thrush, C.R. (2013), "Relationships between the Survey of Organizational Research Climate (SORC) and self-reported research practices", *Sci Eng Ethics.*;19(3): p.835-50, https://doi.org/10.1007/s11948-012-9409-0

Dahl, R.A. (2006), "On political equality", Yale University Press

Dahler-Larsen, P. (2014), "Constitutive Effects of Performance Indicators: Getting beyond unintended consequences", *Public Management Review*, 16:7, p.969-986, https://doi.org/10.1080/14719037.2013.770058

Dahler-Larsen, P. (2019), "Quality – From Plato to Performance", *Palgrave Macmillan*, https://doi.org/10.1007/978-3-030-10392-7

De Vries, R., Anderson, M.S., Martinson, B.C. (2006), "Normal Misbehavior: Scientists Talk About the Ethics of Research", *J Empir Res Hum Res Ethics.*; 1(1): p.43-50, https://doi.org/10.1525/jer.2006.1.1.43

Desrosières, A. (1998), "The Politics of Large Numbers – A History of Statistical Reasoning", Harvard University Press, ISBN 9780674009691

Eitel, B., Mlynek, J. (eds.; 2014), "Führen(d) in der Wissenschaft – Sind Erfolge in der Wissenschaft auch eine Frage der Führung?", *Hanns Martin Schleyr-Stiftung*, Band 85

Fochler, M. & De Rijcke, S. (2017), "Implicated in the Indicator Game? An Experimental Debate", *Engaging Science, Technology, and Society* 3, p.21-40, https://doi.org/10.17351/ests2017.108

Fohlmeister J., & Helling Ch. (2012), "Career situation of female astronomers in Germany", *Astron. Nachr.* / AN 333, No. 3, p.280-286, https://doi.org/10.1002/asna.20121165

Foroutan, N. (2019), "Die postmigrantische Gesellschaft: ein Versprechen der pluralen Demokratie" transcript Verlag

Frese, M., Keith, N. (2015), "Action Errors, Error Management, and Learning in Organizations", *Annual Review of Psychology*, https://www.annualreviews.org/doi/10.1146/annurev-psych-010814-015205

Fricke, W. (2014), "Aktionsforschung in schwierigen Zeiten", In: M. Jostmeier, A. Georg & H. Jacobsen (eds.), "Sozialen Wandel gestalten" (p.213-236). Wiesbaden: Springer Fachmedien

Gagné, M.; Forest, J.; Gilbert, M.-H.; Aubé, C.; Morin, E.; & Malorni, A. (2010), "The Motivation at Work Scale: Validation evidence in two languages", *Educational and Psychological Measurement* 70, p.628–646, https://doi.org/10.1177/0013164409355698.

Gagné, M.; Forest, J.; Vansteenkiste, M.; Crevier-Braud, L.; van den Broeck, A.; Aspeli, A.K.; Bellerose, J.; Benabou, C.; Chemolli, E.; Güntert, S.T.; Halvari, H.; Laksmi Indiyastuti, D.; Johnson, P.A.; Hauan Molstad, M.; Naudin, M.; Ndao, A.; Hagen Olafsen, A.; Roussel, P.; Wang, Z.; Westbye, C. (2015), "The Multidimensional Work Motivation Scale: Validation evidence in seven languages and nine countries, European Journal of Work and Organizational Psychology", 24:2, p.178-196, https://doi.org/10.1080/1359432X.2013.877892

Grecu, V., & Ipiña, N. (2014), "The sustainable University- A Model for the Sustainable Organization", *Management of Sustainable Development,* Sibiu, Romania, Volume 6, No.2, p.15-24, https://doi.org/10.1515/msd-2015-0002

Habermas, J. (1968): Stichworte zu einer Theorie der Sozialisation. In: J. Habermas (1973) *ders., Kultur und Kritik* (p.118-194). Suhrkamp.

Haken, H., Schiepek, G. (2006, [2010]): "Synergetik in der Psychologie – Selbstorganisation verstehen und gestalten", Hogrefe

Havemann, F., Birger, L. (2015), "Bibliometric indicators of young authors in astrophysics: Can later stars be predicted?" *Scientometrics*, 102, p.1413–1434, https://doi.org/10.1007/s11192-014-1476-3

Haven, T., van Woudenberg, R. (2021), "Explanations of Research Misconduct, and How They Hang Together", *Journal for General Philosophy of Science* 52: p.543-561, https://doi.org/10.1007/s10838-021-09555-5

Haven, T.L. (2021), "Towards a responsible research climate: findings from academic research in Amsterdam", https://research.vu.nl/en/publications/towards-a-responsible-research-climate-findings-from-academic-res

Haven, T.L., Bouter, L.M., Smulders, Y.M., Tijdink, J.K. (2019b), "Perceived publication pressure in Amsterdam – survey of all disciplinary fields and academic ranks", *PLoS One;14(6)*, https://doi.org/10.1371/journal.pone.0217931

Haven, T.L., Tijdink, J.K., Martinson, B.C., Bouter, L.M (2019a), "Perceptions of research integrity climate differ between academic ranks and disciplinary fields: Results from a survey among academic researchers in Amsterdam", *PLoS ONE* 14(1): e0210599, https://doi.org/10.1371/journal.pone.0210599

Hayes, D.P. (1992), "The growing inaccessibility of science", *Nature*, Vol 356, p.739-740

Heuritsch, J. (2018), "Insights into the effects of indicators on knowledge production in Astronomy", https://arxiv.org/abs/1801.08033

Heuritsch, J. (2019), "Effects of metrics in research evaluation on knowledge production in astronomy A case study on Evaluation Gap and Constitutive Effects", In: Proceedings of the STS Conference Graz 2019, Graz, Austria, 6-7 May 2019; https://doi.org/10.3217/978-3-85125-668-0-09.

Heuritsch, J. (2021a), "The Evaluation Gap in Astronomy - Explained through a Rational Choice Framework", *under review*; pre-print on ArXiv: https://arxiv.org/abs/2101.03068

Heuritsch, J. (2021b), "Reflexive Behaviour: How Publication Pressure Affects Research Quality in Astronomy", *Publications*, p. 9-52, https://doi.org/10.3390/publications9040052

Heuritsch, J. (2021c), "Reflecting on Motivations: How Reasons to Publish affect Research Behaviour in Astronomy", *under review*; pre-print on ArXiv: https://arxiv.org/abs/2111.15532

Hicks, D., Wouters, P., Waltman, L., De Rijcke, S., & Rafols, I. (2015), "The Leiden Manifesto for research metrics". *Nature*, 520, p.429–431. https://doi.org/10.1038/520429a

John L.K., Loewenstein G., Prelec D. (2012) "Measuring the Prevalence of Questionable Research Practices With Incentives for Truth Telling", *Psychological Science*; 23(5). p.524-532, https://doi.org/10.1177/0956797611430953

Johnston, K.V. (2019), "A dynamical systems description of privilege, power and leadership in academia", *Nat Astron* 3, p.1060-1066, https://doi.org/10.1038/s41550-019-0961-2

Kristiansen, M., Bloch-Poulsen, J. (2005): "Midwifery and Dialogue in Organizations", *Emergent, Mutual Involvement in Action Research*. München und Mering

Kurtz, M. J. & Henneken, E. A., (2017), "Measuring Metrics - A 40-Year Longitudinal Cross-Validation of Citations, Downloads, and Peer Review in Astrophysics", *Journal of the Association for Information Science and Technology*, p. 695-708, https://doi.org/10.1002/asi.23689

Loeb, A. (2010), "The right kind of risk", *Nature*, Vol 467: p.358.

Loeb, A. (2013), "Thinking outside the simulation box", *Nature Physics*, Vol 9: p. 384-386.

Loeb, A. (2014), "Benefits of diversity", *Nature*, Vol 10: p. 616-617.

Lorenz, C. (2012), "If You're So Smart, Why Are You under Surveillance? Universities, Neoliberalism, and New Public Management", *Critical Inquiry*; 38(3), p.599-629, https://doi.org/10.1086/664553

Martinson, B.C., Anderson, M.S., Crain, A.L., De Vries, R. (2006), "Scientists' perceptions of organizational justice and self-reported misbehaviors.", *J Empir Res Hum Res Ethics*; 1(1): p.51-66, https://doi.org/10.1525/jer.2006.1.1.51

Martinson, B.C., Anderson, M.S., de Vries, R. (2005) "Scientists behaving badly", *Nature*; 435(7043), p.737-738, https://doi.org/10.1038/435737a

Martinson, B.C., Crain, A.L., Anderson, M.S., De Vries, R. (2009), "Institutions' Expectations for Researchers' Self-Funding, Federal Grant Holding, and Private Industry Involvement: Manifold Drivers of Self-Interest and Researcher Behavior", *Academic Medicine*; 84(11): p.1491-1499, https://doi.org/10.1097/ACM.0b013e3181bb2ca6

Martinson, B.C., Crain, L.A., De Vries. R., Anderson, M.S. (2010), "The importance of organizational justice in ensuring research integrity", *J Empir Res Hum Res Ethics;5(3):* p.67-83, https://doi.org/10.1525/jer.2010.5.3.67

Martinson, B.C., Nelson, D., Hagel-Campbell, E., Mohr, D., Charns, M.P., Bangerter, A. (2016), "Initial results from the Survey of Organizational Research Climates (SOuRCe) in the U.S. department of veterans affairs healthcare system", *PLoS One.;11(3)*: p.1-18, https://doi.org/10.1371/journal.pone.0151571

Martinson, B.C., Thrush, C.R., Crain, A.L. (2013), "Development and validation of the Survey of Organizational Research Climate (SORC)", *Sci Eng Ethics.;19(3)*: p.813-834, https://doi.org/10.1007/s11948-012-9410-7

Mayring, P. (2000), "Qualitative Content Analysis, Forum: Qualitative Social Research (ISSN 1438-5627), Volume 1, No. 2, Art. 20

Miyakawa, T., (2020), "No raw data, no science: another possible source of the reproducibility crisis", Molecular Brain, https://doi.org/10.1186/s13041-020-0552-2

Moosa, I.A. (2018); "Publish or Perish – Perceived Benefits versus Unintended Consequences", *Edward Elgar Publishing*, https://doi.org/10.4337/9781786434937

Morgan, G. (2018), "Bilder der Organisation", Sonderausgabe. Schäffer-Poeschel

Munafò, M., Nosek, B., Bishop, D., Button, K., Chambers, C.D., Percie du Sert, N., Simonsohn, U., Wagenmakers, E.J., Ware, J., Ioannidis, J.P.A. (2017), "A manifesto for reproducible science". *Nat Hum Behav* 1, 0021, https://doi.org/10.1038/s41562-016-0021, https://www.nature.com/articles/s41562-016-0021.

Parsons, B., (2009), "Evaluative Inquiry for Complex Times", *OD Practioner*, Vol:41, No.1, p.44-49, https://insites.org/resource/evaluative-inquiry-for-complex-times/

Patat, F., Boffin, H.M.J., Bordelon, D., Grothkopf, U., Meakins, S., Mieske, S. & Rejkuba, M., (2017), "The ESO Survey of Non-Publishing Programmes", *European Southern Observatory*, p. 51-57, https://doi.org/10.18727/0722-6691/5055

Pateman, C. (1970), "Participation and Democratic Theory", Cambridge University Press

Pluchino, A., Biondo, A. E., Rapisarda, A., (2018), "Talent vs Luck: the role of randomness in success and failure", *Advances in Complex Systems*, Vol. 21, No. 03n04, 1850014, https://doi.org/10.1142/S0219525918500145

Schnell, T. (2018), "Von Lebenssinn und Sinn in der Arbeit. Warum es sich bei beruflicher Sinnerfüllung nicht um ein nettes Extra handelt", In: Fehlzeiten-Report, p.11-21. Berlin: Springer.

Sunstein, C.R. (2018), "# Republic: Divided democracy in the age of social media", Princeton University Press

Tijdink, J.K., Verbeke, R., Smulders, Y.M. (2014), "Publication pressure and scientific misconduct in medical scientists", *J Empir Res Hum Res Ethics*; 9(5): p.64-7, https://doi.org/10.1177/1556264614552421

Van Dyck, C., Frese, M., Baer, M., Sonnentag, S. (2005), "Organizational Error Management Culture and Its Impact on Performance: A Two-Study Replication", *Journal of Applied Psychology*, https://doi.apa.org/doiLanding?doi=10.1037%2F0021-9010.90.6.1228

Wellcome Trust (2020), "What Researchers Think About the Culture They Work In", https://wellcome.org/reports/what-researchers-think-about-research-culture

Wells, J.A., Thrush, C.R., Martinson, B.C., May, T.A., Stickler, M., Callahan, E.C. (2014), "Survey of organizational research climates in three research intensive, doctoral granting universities", *J Empir Res Hum Res Ethics;9(5)*: p.72-88, https://doi.org/10.1177/1556264614552798

Zuiderwijk, A. & Spiers, H., (2019), "Sharing and re-using open data: a case study of motivations in astrophysics", *International Journal of Information Managment*, Elsevier Ltd, p. 228-241, https://doi.org/10.1016/j.ijinfomgt.2019.05.024